\documentclass[5p,twocolumn,times]{elsarticle}

\usepackage{graphicx,isolatin1}
\usepackage[squaren]{SIunits}

 \usepackage{etoolbox}

\newcommand{\textroman}[1]{\mbox{\rm #1}}
\newcommand{\gfrac}[2]{\displaystyle\frac{#1}{#2}}

\newcommand{\dd}{\mbox{d}}

\RequirePackage{xspace}
\def\calA{{\ensuremath{\cal A}}\xspace}

\def\textE{{\ensuremath{\textroman{E}}}\xspace}
\def\eff{\textroman{eff}}

\journal{Nucl.\ Instrum.\ Meth.\ A }

\begin{document}

\begin{frontmatter}

\title{Polarimetry of cosmic gamma-ray sources above $e^+e^-$ pair
  creation threshold }

\author[add1]{D.~Bernard\corref{cor}}
\ead{denis.bernard at in2p3.fr}

\address[add1]{LLR, Ecole Polytechnique, CNRS/IN2P3,
91128 Palaiseau, France}

\cortext[cor]{Tel 33 1 6933 5534}

\begin{abstract}
We examine the potential for gamma-ray conversion to electron-positron pairs,
either in the field of a nucleus or of an electron of a detector, to
measure the fraction $P$ of linear polarization of cosmic gamma sources.
For this purpose 
we implement, validate and use an event generator based on the HELAS
amplitude calculator and on the SPRING event generator.

We  characterize several ways to measure $P$.
Past proposals to increase the polarization sensitivity by the
selection of a fraction of the events in a subset of the available
phase space are found to be inefficient, due to the loss in
statistics.
The use of an optimal variable that includes the full 5D probability
density function is found to improve the precision of the measurement
of $P$ of a factor of approximately 2.

We then study the dilution of the asymmetry that parametrize the
degradation of the precision due to experimental effects such as
multiple scattering.
In a detector made with a succession of converter slabs and tracker
foils, the dependence of the dilution  is
found to be different from that predicted assuming a given (the most
probable) value of the pair opening angle.
The limitations of a slab detector are avoided by the use of an active
target, in which conversion and tracking are performed by the same
device, in which case the dilution of the measurement of $P$ is found
to be manageable.
Based on a realistic sizing of the detector, and for an effective
exposure of 1 year, we estimate the precision for a Crab-like source
on the full energy range to be approximately 1.4\,\%.
\end{abstract}


\end{frontmatter}

\newtoggle{couleur}
\toggletrue{couleur}

In section \ref{sec:astro} we present a physics case for $\gamma$
polarimetry in the MeV -- GeV photon energy range.
Section \ref{sec:polarimetry:how:to} is an introduction to the
measurement technique, focused on pair conversion.
In section \ref{sec:evt:gen} various aspects of the measurement are
studied at generator level, that is in absence of any experimental
effect.
The generator is validated by comparison with published results based on
analytical calculations. The distributions of several kinematic
variables of interest, such as the opening angle and the transferred
momentum, are studied.
Some differences between nuclear and triplet conversion are pointed out.
We examine several past proposals to improve the $P$ sensitivity,
after which we apply the technique of ``optimal variables'' to make
use of all the information present in the 5D probability density
function (pdf).
We then study the effects of multiple scattering and of experimental cuts
in section \ref{sec:expt:effects}.
Finally we describe the performance of a realistic detector, a
$1\,\meter^3$ $5\,\bbar$ argon gas TPC.

\section{A science case for $\gamma$ Polarimetry} 
\label{sec:astro}

In many sources of gamma rays, the models proposed to explain the 
emission have very different polarization signatures. Depending on the
orientation of the magnetic field at the source and on the primary
emission mechanism, different degrees and directions of polarization
are anticipated. $\gamma$-ray polarimetry can therefore be used to
probe the nature and geometry of many objects including pulsars,
binary systems, gamma-ray bursts and active galactic nuclei.

In {\bf rotation powered pulsars} (RPPs), the pulsed, non-thermal
radiation from relativistic particles in the magnetosphere of the
neutron star is highly beamed along its magnetic field lines. This
means that the emitted radiation should be highly polarized either
parallel or perpendicular to the field lines
\cite{Kaspi:2004:IsolNeutronStars}. Although the different high-energy
emission models share common emission mechanisms, these processes
occur at different locations in the pulsar system depending on the
geometry of the particular emission model. Thus, the polarization
signatures of these models are expected to differ greatly from each
other. 
Reference
\cite{Slowikowska:2009} shows the different polarization
signatures expected at optical wavelengths. Similar degrees of
polarization are expected in the keV - MeV energy range but with a
different polarization angle \cite{McConnell:2009}. This
energy-dependent rotation of the polarization direction could be used
to locate the sites of emission and to probe the emitting particle
population and their energetics \cite{McConnell:2009}. 

Polarized high-energy emission has also been detected from the {\bf Crab
pulsar wind nebula} (PWN) between 200\,keV and 800\,keV 
\cite{Forot:2008} and between 100\,keV and 1\,MeV
 \cite{Dean:2008}. In both cases the polarization was aligned with
the spin axis of the neutron star.

Other galactic systems are also predicted to emit polarized gamma
rays. Many models for the X-ray emission from {\bf accretion powered
pulsars} (APPs) predict high degrees of linear polarization which
varies both with pulse phase and with energy
\cite{McConnell:2009}. Using the calculations of
Ref. \cite{Meszaros:1988}, Ref. \cite{Ghosh:2013} discusses how the polarization
of the emission from these objects up to $\approx$85\,keV can be used
to decipher the long-standing problem of the geometry of the emitting
regions and also to search for the signature of vacuum resonance 
(photon propagation through the birefringent strongly magnetized
plasma leads to polarized emission, but for a critical value of the
field, the birefringence of the vacuum cancels that of the plasma,
leading to resonant patterns in the opacities).
Reference \cite{Dovciak:2008} shows how observations of polarization
in the X-ray regime can be used to test strong gravity in galactic
{\bf black hole binaries}.

{\bf GRB} fall broadly into two categories that are postulated to be
those created by the explosion of a hypernova or as a result of the
coalescence of two compact objects (e.g., neutron stars, white dwarfs,
black holes).
In the relativistic jets that are thus produced,
synchrotron emission is thought to be the dominant emission
process. Electrons are accelerated to near light speed by the
relativistic shocks and, given the presence of a strong magnetic field
\cite{Piran:1999:GRB-Fireballs}, the degree of polarization of the
photons emitted by these jets is expected to be very high.
Polarization measurements of the emission should therefore provide
information critical to distinguishing between the many emission
models that exist for GRBs (see, for example, Refs. 
\cite{Waxman:2003:PolGRBs, Granot:2003:PolGRBs,Eichler:2003:PolGRBs,
 Lyutikov:2003:PolGRBs}). Further Compton scattering of synchrotron
photons on cold electrons is expected to decrease the polarization
fraction and to rotate its angle by 90\degree\ at high energy, so 
polarimetry in energy bins is needed \cite{Chang:2013xp}.

Another source class whose study should benefit from polarimetry
studies is {\bf AGN} and, 
in particular, the {\bf blazar}
subclass, which are strong emitters in the $\gamma$ energy range.
The models proposed to explain their emission fall into two broad
categories, namely, hadronic and leptonic.
One study of the polarization properties of relativistic jets
\cite{McNamara:2009:XrayPolJets} has shown that the observed degree of
X-ray polarization should be sufficiently different for two of the
most commonly postulated leptonic emission mechanisms,
external-Compton (EC) and synchrotron-self-Compton (SSC), to determine
which (if either) is the dominant process in blazars.
Analytical calculations and simulations show that in contrast with EC
emission, which is expected to be polarized at a level of $\ll 1 \%$,
SSC emission of a blazar can be polarized to more than 50 \% and that
the degree of polarization rises above the MeV range, while it
plateaus at X-rays energies \cite{Krawczynski:2011ym}.
If it can be shown that hadrons are accelerated to high energies in
the jets of AGN, this source class would be a strong candidate for the
accelerators of the high-energy cosmic rays, a long-standing mystery
in astrophysics.

In the course of the efforts to build a quantized theory of
gravitation, the possibility of
 {\bf Lorentz and CPT invariance violation} has been
considered.
$\gamma$ polarimetry turns out to be the most sensitive tool to test
such an effect.
In the framework of effective field theories, a birefringence effect of
the vacuum is predicted \cite{LIV}.
The
linear polarization direction would be rotated through an energy-dependent
angle, due to different phase velocities for opposite helicities. This
vacuum birefringence would rotate the polarization direction of
monochromatic radiation, or could depolarize linearly polarized
radiation composed of a spread of energies.
The rotation angle is expressed as $\theta \approx \xi E^2 t /(2M_P)$
where $E$, $t$, $M_P$ are the photon energy, the propagation time, and
the Planck mass, and $\xi$ is a dimensionless Lorentz-violating
effective field theory (EFT) parameter. 
The present limit is of $| \xi | < 3.4 \times 10^{-16}$
\cite{Gotz:2013dwa}
based on the observation of the polarization of the X-soft-$\gamma$
emission of GRB 061122 in the 250 -- 800 keV energy range with IBIS on
INTEGRAL.
Due to the squared dependence $\theta \propto E^2$, extending the
polarization measurements to higher energies would lead to an improved
sensitivity to a possible Lorentz invariance violation.

Last but not least, GRB polarimetry allows us to search for hints of the
presence of the {\bf axion}, the pseudo-scalar field associated with
the $U(1)$ symmetry devised to solve the QCD CP problem.
The dichroism induced by the coupling of the propagating photon with
the axion in the presence of the magnetic field generated by the GRB
would lead to a rotation of the polarization direction which is here
proportional to 
the photon energy : again, due to the width of the energy spectrum the polarization would be blurred.
The actual observation of a non-zero polarization fraction would allow us to
obtain an upper limit on the axion-to-two-photon coupling
$g_{a\gamma\gamma}$, 
which is GRB-model dependent, but it is presently the best limit for an
axion mass close to $1\,\milli\electronvolt$
 \cite{Rubbia:2007hf}.
As the limit is proportional to $1/\sqrt{E}$, extending the
polarization measurement to higher energies would lead to an improved
value, or even to a detection.

\section{$\gamma$-ray polarimetry}
\label{sec:polarimetry:how:to}

The polarization fraction of a gamma-ray beam is measured by analyzing 
 the distribution of the azimuthal angle, $\phi$,
of its conversion in a detector, which is given by
the differential interaction rate
\begin{equation}
\frac{\dd \Gamma}{\dd \phi} \propto 
(1 + \calA P \cos{[2(\phi-\phi_0)]}),
\label{eq:diff1D}
\end{equation}

where $P$ is the fraction of the linear polarization of the photon beam and
$\calA$ is the polarization asymmetry of the conversion process.
There are several ways to define the azimuthal angle, $\phi$, as we
shall see later.
The angle origin, $\phi_0$, determines the orientation of polarization
of the photon flux from a given cosmic source, with respect to a given
fixed direction.

\subsection{Compton scattering}

In the case of Compton scattering, 
$\calA \approx 2 m/E$ at high photon energy, $E$, 
 so the sensitivity to polarization strongly decreases above a
 few MeV \cite{McConnell-Ryan}.

\subsection{Pair conversion}

\begin{figure}[htp]
\centerline{\includegraphics[width=0.69\linewidth]{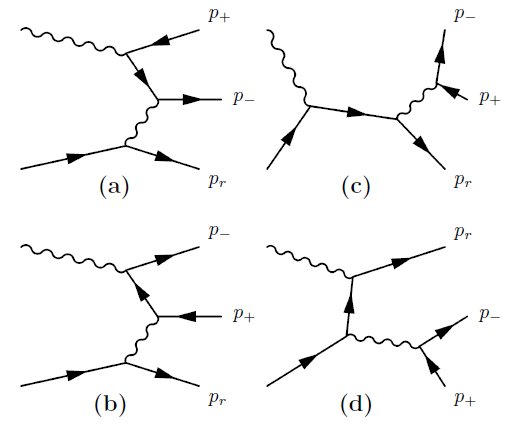}}
\caption{Feynman diagrams describing photon conversion in the Born
 approximation; (a) and (b) are called Borsellino diagrams, and (c)
 and (d) are called $\gamma-e$ diagrams.
Particles are noted $+$ (positron), $-$ (electron) and $r$ (recoil),
and labeled by their momentum, $p$.
\label{fig:Feynman}
}
\end{figure}

A photon with energy above the pair creation threshold can convert to
an $e^+e^-$ pair in the electric field of a charged particle of the
detector.
The process is described in the Born approximation by the diagrams
shown in Fig. \ref{fig:Feynman}.

\subsubsection{Nuclear pair conversion}

``Nuclear'' conversion, in the field of a nucleus, was
first considered for polarimetry in 1950 \cite{ber}.
This process dominates the cross-section at high energy where $\calA$
goes to a constant.
In that case, $\phi$ is measured from the $e^+e^-$ pair.

It was soon realized, however, that multiple scattering of the tracks
blurs the measurement of $\phi$, which induces a damping of the
modulation in eq. (\ref{eq:diff1D}).
The asymmetry is reduced and an effective asymmetry $\calA_{\eff}$ can
be written as $\calA_{\eff} = D \times \calA$, where $D$ is a dilution
factor, $D = e^{-2 \sigma_{\phi}^2}$ \cite{Kotov,Mattox,Bloser:2003gb}, 
and $\sigma_{\phi}$ is the
$\phi$ angle resolution, which after the propagation over a length $L$
of material with radiation length $X_0 $, is $\sigma_{\phi} \approx 14
\sqrt{L/X_0}$ \cite{Kotov,Mattox}.
A dilution of $D = 1/2$ is obtained for a resolution of
$\sigma_{\phi} = \sigma_c \equiv \sqrt{\ln{2}/2} \approx 0.59\,\radian$.
This limitation is energy independent because, despite the fact that
the electrons from higher-energy photon conversions have larger
momenta and therefore suffer less multiple scattering, the pair is
emitted in the forward direction with a typical opening angle that
decreases with increasing photon energy as $m/E$, where $m$ denotes
the electron mass.
The direction of recoil of the nucleus also keeps track of the
polarization of the photon, but the recoil momentum is of the order
$\mega\electronvolt/c$ and the path length of the nucleus is too
short to be detected and therefore to enable a measurement of $\phi$.

Using this approach, a $\gamma$-ray polarimeter using nuclear
conversion with a detector comprised of slabs would require a huge
number of extremely thin converters, which was considered to be
infeasible.

\subsubsection{Triplet conversion}

When
 the incoming photon converts in the field of
an electron, $\gamma e^- \rightarrow e^- e^+ e^- $, 
the recoiling electron is emitted at a large angle with respect to the
photon direction, and therefore the measurement of $\phi$ is easier.
Three tracks are observed in the final state, and thus the process
is named ``triplet'' conversion \cite{Perrin1933}.
Since two electrons are present in the final state, four additional
``exchange'' diagrams are present, which are not shown in
Fig. \ref{fig:Feynman}.
Votruba first derived an expression, albeit a very
tedious one, for the total cross-section taking into account all eight
of the diagrams \cite{Votruba1948}.
Borsellino subsequently showed that diagrams (a) and (b) dominate the
triplet cross-section at high energy and that the $\gamma-e$
diagrams and the exchange diagrams could therefore be neglected.
Most of the subsequent works (e.g. Refs. \cite{SuhBethe1959},
\cite{Endo:1992nq}) studied the energy range above 50\,MeV, 
which is unfortunate as for cosmics sources most of the signal is
below that value.

\subsubsection{Sign of the asymmetry}

It has been shown \cite{Maximon:1962zz}
that in the case of nuclear conversion, the plane of
emitted $e^+e^-$ pair correlates with the direction of the
 polarization of the photon in such a way that the pair is preferably
emitted in the plane of polarization of the photon ($\calA >0$) 
\footnote{Except for the events for which the pair is almost coplanar
 to the photon direction \cite{Maximon:1962zz}, for which the asymmetry changes sign, which is confirmed by our simulation.},
while for triplet conversion the recoil electron is preferably emitted
in the plane orthogonal to the direction of the photon polarization 
 ($\calA <0$) \cite{Boldyshev:1972va,Boldyshev:1994bs}.
The simulation developed here confirms the theoretical calculations
discussed in Ref. \cite{Maximon:1962zz}. 

Obviously the point here is not the charge of the recoiling particle,
as changing its sign would be simply equivalent to a flip of the
photon electric field (e.g. from $x$ to $-x$) that does not affect the
polarization.
Simply, 
 an azimuthal angle determined from the plane of the pair
yields a positive $\calA$, while an azimuthal angle determined from
an (approximately) normal to that plane (such as the pair momentum or
the recoiling particle direction) yields a negative $\calA$.
Needless to say, exchanging by mistake the electron and the positron
of the pair does not affect the measurement.

Asymmetries have been measured in two experiments, both with nuclear
conversion and at high energy ($\approx \giga\electronvolt$);
strict acoplanarity was not required \cite{Adamyan,deJager}.
The sign of the asymmetry is not discussed explicitly in either of
these works.
Reference \cite{Adamyan} seems to obtain a positive $\calA$, and so is
Ref. \cite{deJager} (See their Fig. 6) when an offset of 
$\phi_0 = \pi/2$ is taken into account\footnote{Bogdan Wojtsekhowski,
private communication, May 2013.}.
In the following we do not consider this issue explicitly, and assume
implicitly a positive value for $\calA $, for convenience.

\subsection{Energy range of interest}

\begin{figure}
\iftoggle{couleur}{
\centerline{\includegraphics[width=0.7\linewidth]{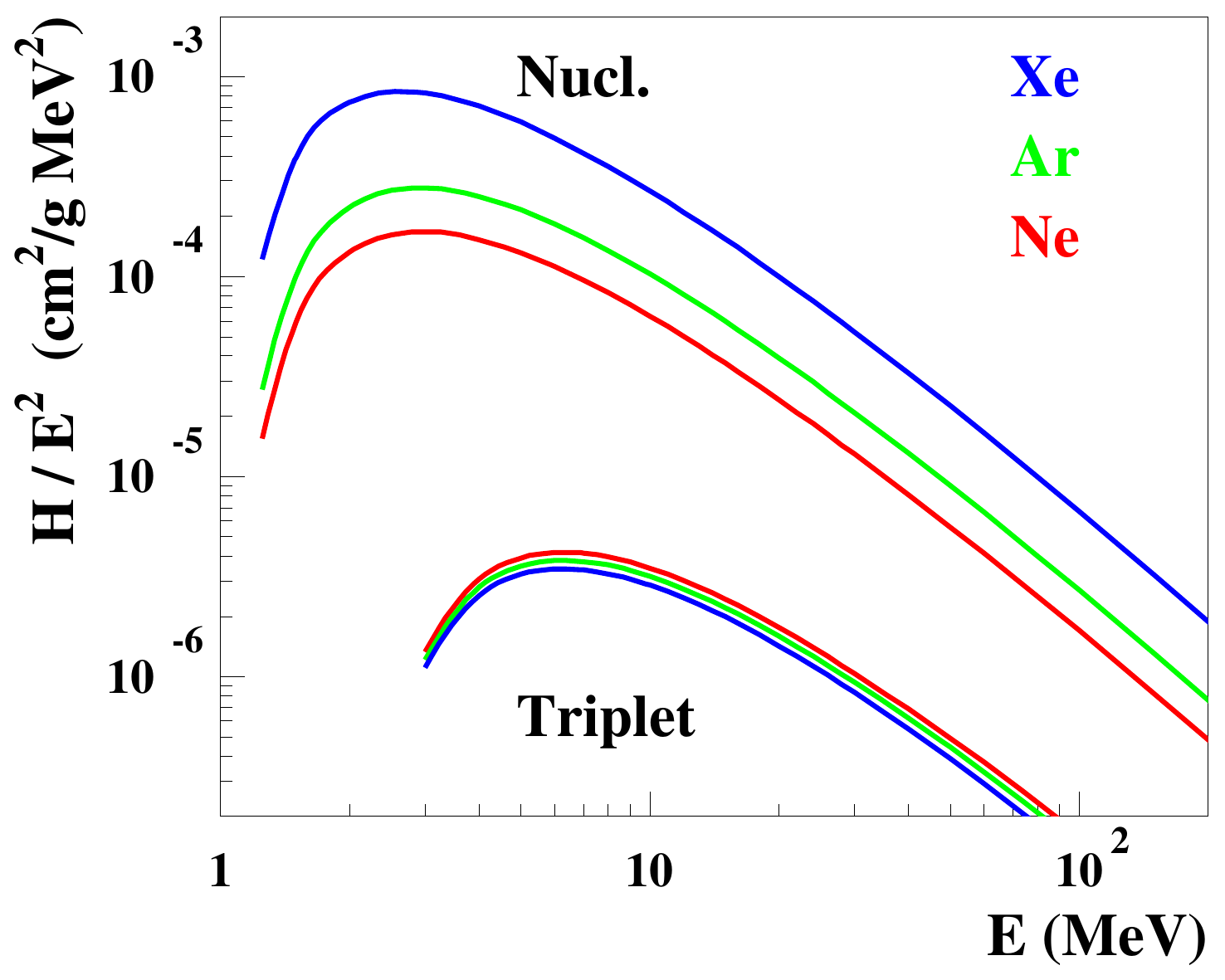}}
}{
}
\caption{Product of the attenuation length for three noble gases by 
a typical cosmic-source spectrum $1/E^2$ as a function of photon energy $E$
($H$ is taken from Ref. \cite{NIST}).
\label{fig:energy-range}
}
\end{figure}

The probability of the conversion of a photon in a detector is
determined by the interaction length $\Lambda = 1/(H \rho)$, where $H$
is the mass attenuation coefficient in the detector material and
$\rho$ is the material density.
The value of $H$ is tabulated by NIST \cite{NIST} as a function of
photon energy. It rises rapidly 
above threshold, which is $2 m c^2$
for nuclear conversion and $4 m c^2$ for triplet.
Most cosmic $\gamma$-ray sources have a power-law spectrum
that decreases with energy, such that the flux, $F \propto
E^{-\Gamma}$, where $\Gamma$ is the spectral index, which typically
has a value of $\approx 2$.
The energy range of interest is obtained by examining the variation
with $E$ of the product $F \times H$ (Fig. \ref{fig:energy-range}).
Most of the conversions and therefore most of the sensitivity to
polarization clearly lie below 50\,MeV.

The high-energy asymptotic differential cross-section, which is valid
both for triplet and nuclear conversion (after $Z^2$ scaling) is
\cite{Boldyshev:1972va}:
\begin{center} \small
\begin{equation} 
\frac{\dd \sigma} {\dd \phi}
\propto
\alpha r^2_0
\left(
\left[\frac{28}{9}\ln{2\omega}-\frac{218}{27}\right]
- P \cos{[2(\phi-\phi_0)]}
\left[\frac{4}{9}\ln{2\omega}-\frac{20}{27}\right]
\right),
\end{equation}
\end{center}

with $\omega \equiv E/m$. 
The fine-structure constant is denoted $\alpha$ and the classical
radius of the electron, $r_0$.
This expression leads to an asymptotic value for
$\calA$ of $1/7 \approx 14 \%$ and undergoes an unphysical change of sign
for $\ln{2\omega} = 5/3$, that is for $E=1.35\,\mega\electronvolt$.
Corrections of order $m/E$ to $\calA$ have been computed in Ref. 
\cite{Akushevich:1999kw}, 
but the same behavior remains, with $\calA$ decreasing and changing 
sign at low energy.

\section{The event generator}
\label{sec:evt:gen}

This work builds on the previous opus by Endo and 
Kobayashi \cite{Endo:1992nq}. 
In Ref. \cite{Endo:1992nq} the differential cross-section for triplet
conversion was
computed exactly, including the eight diagrams, within the
Born approximation and without screening.
After the amplitudes were computed using the HELAS software
\cite{Murayama:1992gi},
the differential cross-section was integrated using the BASES
 integrator \cite{Kawabata:1985yt}.
Triplet production was studied in the photon energy range 50 -- 550\,MeV, 
in particular the variation of $\calA$ with the opening angle of the
pair.

Here we use an improved version of BASES \cite{Kawabata:1995th}, 
and we use the event generator SPRING
\cite{Kawabata:1995th}.
We extend the study to nuclear conversion, and we compare the
 HELAS computation to the Bethe-Heitler (BH) differential cross
section \cite{Heitler1954}, an analytical expression based on the two
Borsellino or BH diagrams.
In both cases, the energy that is carried away by the recoiling
particle is taken into account.
The screening of the electric field of the recoiling particle is also
taken into account.

~

\begin{figure}[!h]
 \begin{center}
 \setlength{\unitlength}{0.47\textwidth}
 \begin{picture}(1,0.7)(0,0)
 \put(0,0){
\includegraphics[width=0.47\textwidth]{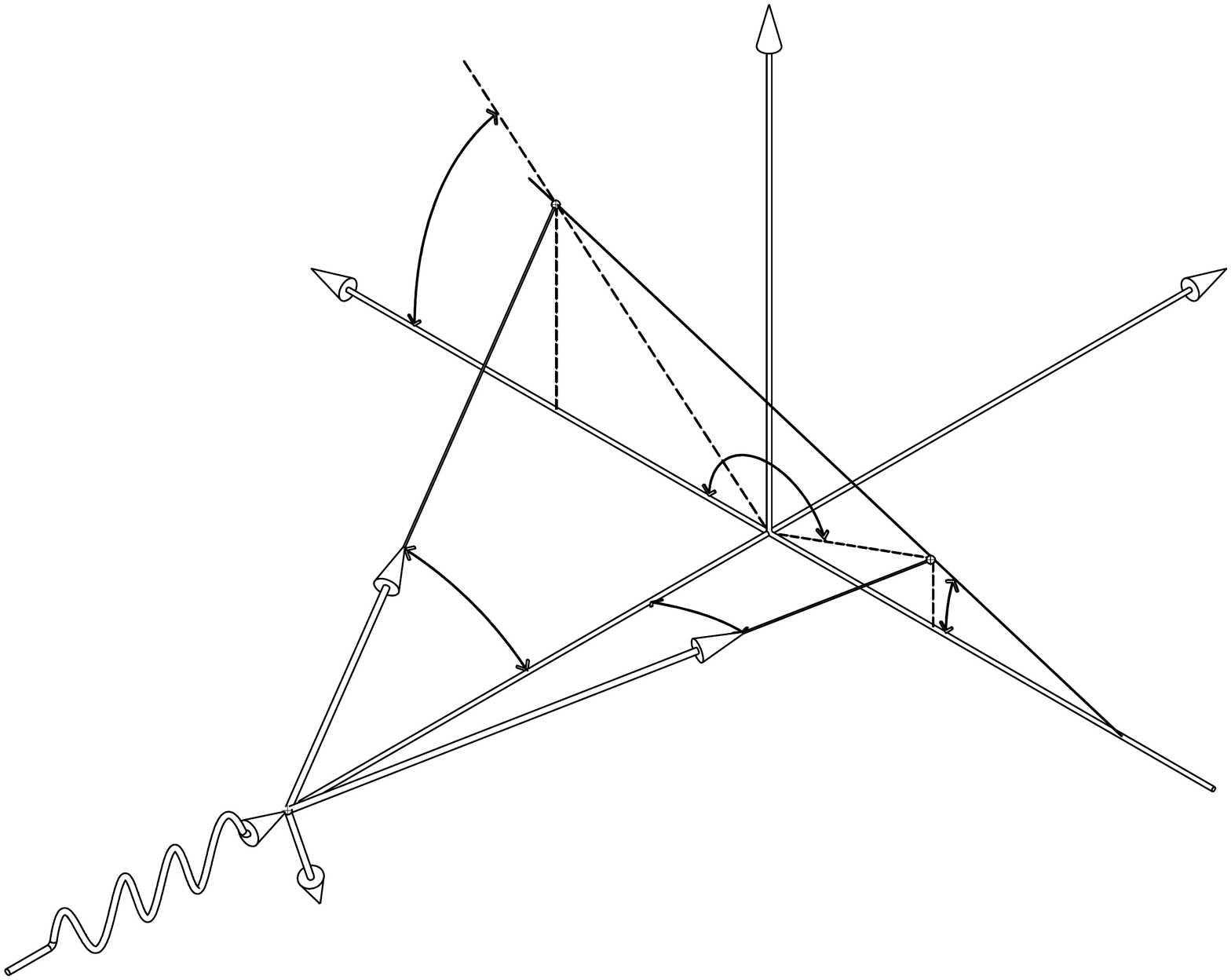}
}
 \put(0.27,0.5){$x$}
 \put(0.6,0.71){$y$}
 \put(0.97,0.5){$z$}
 \put(0.32,0.57){$\phi_+$}
 \put(0.575,0.4){$\phi_-$}
 \put(0.775,0.27){$\omega_{+-}$}
 \put(0.41,0.29){$\theta_+$}
 \put(0.58,0.265){$\theta_-$}
 \put(0.3,0.3){$\vec{p_+}$}
 \put(0.59,0.2){$\vec{p_-}$}
 \put(0.3,0.02){$\vec{p_r}$}
 \put(0.03,-0.01){$\vec{k}$}
 \end{picture}
 \caption{Schema of a photon conversion.
 \label{fig:schema:angles}}
 \end{center}
\end{figure}

The incoming photon is assumed to have a momentum $\vec{k}$ parallel to $z$.
We denote $\vec{p}$, $E$, $\theta$, $\phi$ the momentum, energy, polar and
azimuthal angle of the outcoming particles, with subscripts $+$, $-$,
and $r$ for the positron, electron, and recoiling particle,
respectively
(Fig. \ref{fig:schema:angles}).
The fraction of the photon energy carried away by a lepton is denoted $x$,
e.g. $x_+ \equiv E_+/E$, and
$\vec{q} \equiv \vec{k} - \vec{p_+} - \vec{p_-}$ is the momentum
transferred to the recoiling particle.

With three particles in the final state, and taking energy-momentum
 conservation into account, the final state is described by five
 variables. 
These can be the azimuthal and polar angles describing the two track of
 the pair plus the already mentioned energy fraction that is $\phi_+$,
 $\theta_+$, $\phi_-$, $\theta_-$ and $x_+$,
all in the laboratory frame, as in the expression of the differential
 rate established by Bethe and Heitler \cite{Heitler1954}.
Instead, as in Ref. \cite{Endo:1992nq}, we first generate the
 azimuthal and polar angles of the recoiling particle and of one of
 the track, the positron, all in the center-of-mass frame, and the
 invariant mass of the pair.

\subsection{Validation}

As the 5D differential cross-section is rather involved, with strongly
peaked variables such as the polar angle of the electrons of the pair,
we validate the behavior of the generator against known properties of
the pdf available in the literature.

\subsubsection{cross-section}

We first compare the triplet cross-section obtained by the integrator
of the event simulation
 to the computation by Mork
\cite{Mork1967}, as was done in Ref. \cite{Endo:1992nq}
(no screening). 
This is shown in Fig. \ref{fig:comparemork} (Supplementary data).

In practice the charged particle on which the photon converts is
embedded into an atom of the detector.
When the differential cross-section is computed in the Born
approximation, the screening of the electric field of that particle by
the (other, in the case of triplet conversion) electrons of the atom
can be described by a form factor.
In the case of nuclear conversion, the recoiling nucleus is slow
enough that the collision as regarded by the atom can be considered as
elastic, while in the case of triplet conversion the electron is
ejected and the atom, ionized.

Screening in nuclear conversion is taken into account by multiplying
the differential cross-section by $[1 - F(q)]^2$ where 
$F(q)$ is the Mott atomic form factor, 
$F(q) = 1/(1 + (111 (q/mc) Z^{-1/3})^2)$ 
\cite{Mott:1934}.
We note that screening affects the $q$ spectrum below a typical value 
$q_Z = mc Z^{1/3}/111$ ($\approx 12\,\kilo\electronvolt/c$ for argon).
The $q$ distribution is narrow and centered around 
$1\,\mega\electronvolt/c$ at low energy. 
It becomes extended at high energy (see Fig. \ref{fig:coupe-N}) with
asymptotically a lower kinematic limit of $q_m = 2m^2 /E$
(e.g. \cite{Perrin1933}).
We see that screening affects the lower part of the $q$ spectrum for
photon energies larger than $\approx 2 m^2 /q_Z$ (43\,MeV for argon).
For triplet conversion, screening is taken into account by multiplying
the differential cross-section by the incoherent scattering factor,
$S(\nu)$, where $\nu \propto q$ 
 \cite{WheelerLamb1939}. 
At low $q$ ($\nu < 0.01$) we use $S(\nu) = 13.8 \nu - 55.4 \sqrt{\nu}$ 
 \cite{WheelerLamb1939}. 
At a higher value, $1> \nu > 0.01$, we use 
$S(\nu) = \sqrt{1 - (1-\nu)^2}$, which we found to be a good
representation of the data tabulated in table I of 
Ref. \cite{WheelerLamb1939}.
For $\nu >1$ we use $S(\nu) =1$.

The mass attenuation coefficients computed with the simple Anz\"atze
used here are compared to the full computation from the NIST server
\cite{NIST} in Fig. \ref{fig:formfactor} (Supplementary data, left).
A more precise comparison is presented in Fig. \ref{fig:formfactor}
(Supplementary data, right), where the ratio of the cross-sections with/without screening
is compared to the values tabulated in Table 1 (nuclear) and Table 5
(triplet) of Ref. \cite{Hubbell1980}, for aluminium.
The fair agreement meets the needs of the present study.

At low energy, just above threshold, the interference between the diagrams
in triplet conversion leads to a decrease of the cross-section, from
1.4 to 1.11 $\mu\barn$ at 
$\omega = 4.4$ ($E \approx 2.25\,\mega\electronvolt$) that is
compatible with the analytical computation by Mork 
(1.1 $\mu\barn$ \cite{Mork1967}).
The systematic comparison of the distributions of a series of
kinematic variables, $q$, $\theta_{+-}$, $\theta_{+}$, $x_{+}\equiv
E_{+}/E$ for the BH and H generators do not show any difference
within statistical fluctuations for nuclear conversion (plots not
shown).

\subsubsection{Opening angle}

\begin{figure}
\iftoggle{couleur}{
 \includegraphics[width=0.49\linewidth]{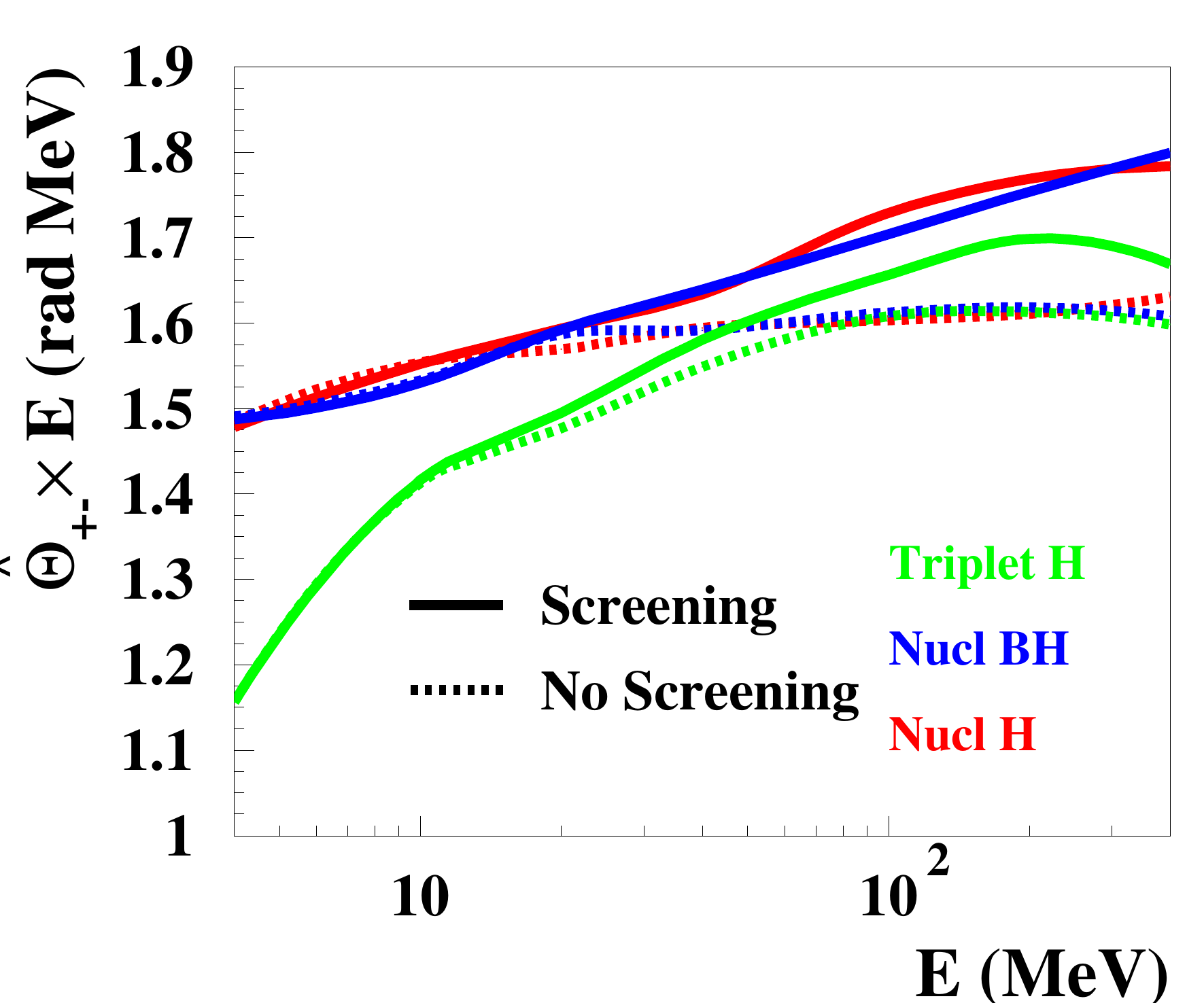}
 \includegraphics[width=0.49\linewidth]{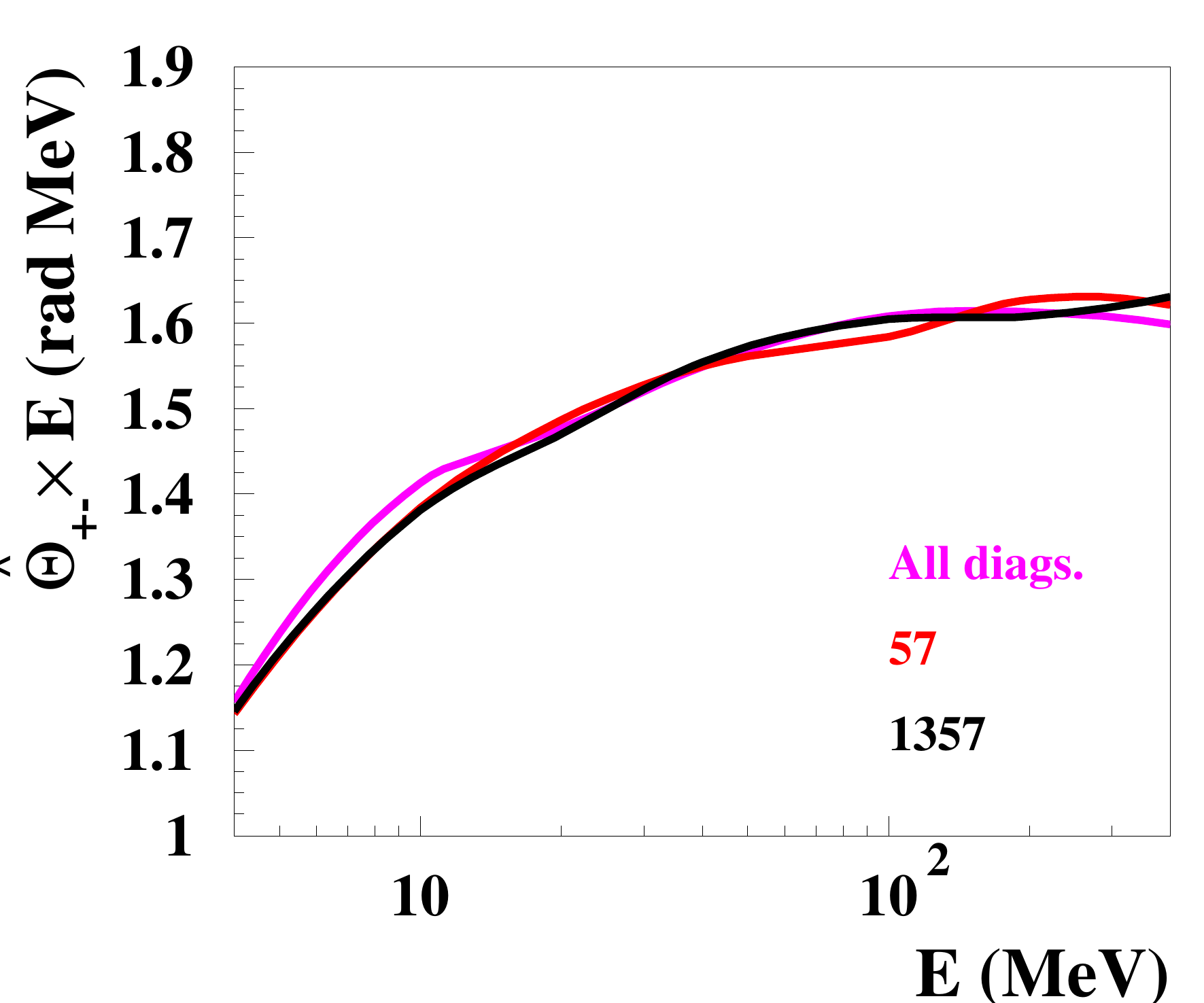}
}{
}
\caption{Position of the maximum $\hat{\theta}_{+-} \times E$ of the
 distribution of $\theta_{+-} \times E$ (\radian\ MeV), as a function
 of photon energy.
Left: for nuclear and triplet, with/without screening, for the H/BH pdf.
Right: triplet with all diagrams, without exchange diagrams (1357) 
and with BH diagrams only (57). 
\label{fig:thetaplusminus}
}
\end{figure}

The distribution of the opening angle $\theta_{+-}$ of the pair 
was computed by Olsen from a high energy approximation of the
differential cross-section, with a most probable value,
$\hat{\theta}_{+-}$, 
that decreases with photon energy as $E_0/E$ with 
$E_0 \approx 1.6\,\mega\electronvolt$ \cite{Olsen1963}.
Experimentally, some differences between the distribution of nuclear
and triplet conversions have been noted \cite{Sandhu}.
We fit the peak of the $\theta_{+-}$ distribution with a third degree polynom, 
we compute $\hat{\theta}_{+-}$ and we present the variation of
$\hat{\theta}_{+-} \times E$ as a function of $E$ 
in Fig. \ref{fig:thetaplusminus} (left).
\begin{itemize}
\item 
For nuclear conversion, the values for the Bethe-Heitler (BH) and the
full (H) amplitude show compatible results.
\item 
When no screening is applied, triplet and nuclear conversion both
show a value compatible with that of Ref. \cite{Olsen1963} at high energy.
\item 
When screening, which is active only at high energy, is taken into
account, $\hat{\theta}_{+-}$ increases and increases with energy.
\item 
For triplet conversion, $\hat{\theta}_{+-}$ decreases strongly at low
energy, which has been observed experimentally \cite{Augerat1962},
with a reduction of almost a factor of two just above threshold.
\end{itemize}

Since
 there are two negative electrons in the final state for triplet
conversion, we choose by convention that the one with the smallest polar
angle ($\theta$) wrt the direction of the incoming photon in the
laboratory frame be the member of the pair.
The other one is considered to be the recoiling particle.
Neglecting the exchange diagrams for triplet conversion, or even
restricting the amplitude to the Borsellino diagrams, does not affect
the value of $\hat{\theta}_{+-}$ (Fig. \ref{fig:thetaplusminus}
right).

\subsubsection{Recoil momentum distribution}

\begin{figure}
\iftoggle{couleur}{
 \includegraphics[width=0.49\linewidth]{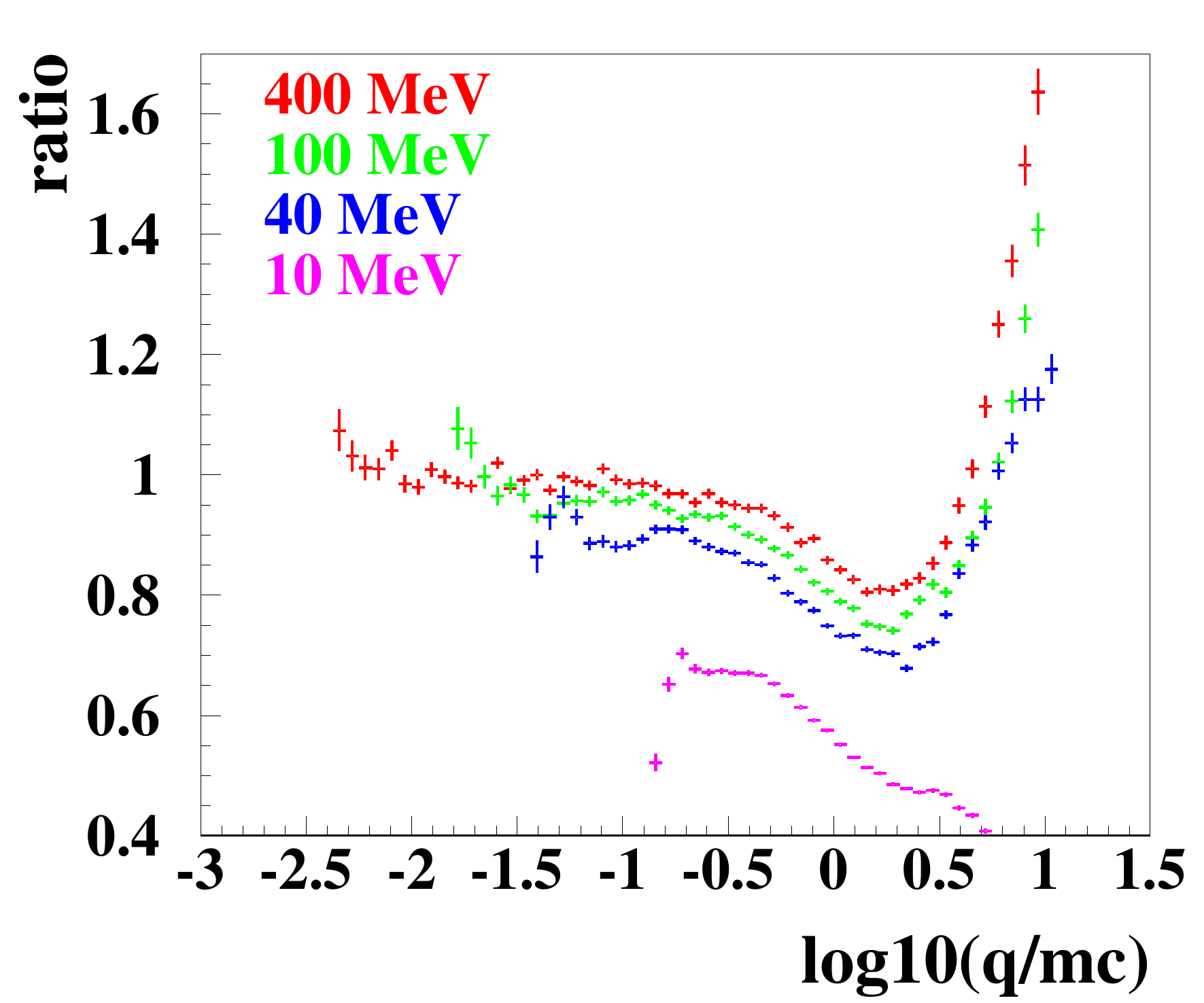}
 \includegraphics[width=0.49\linewidth]{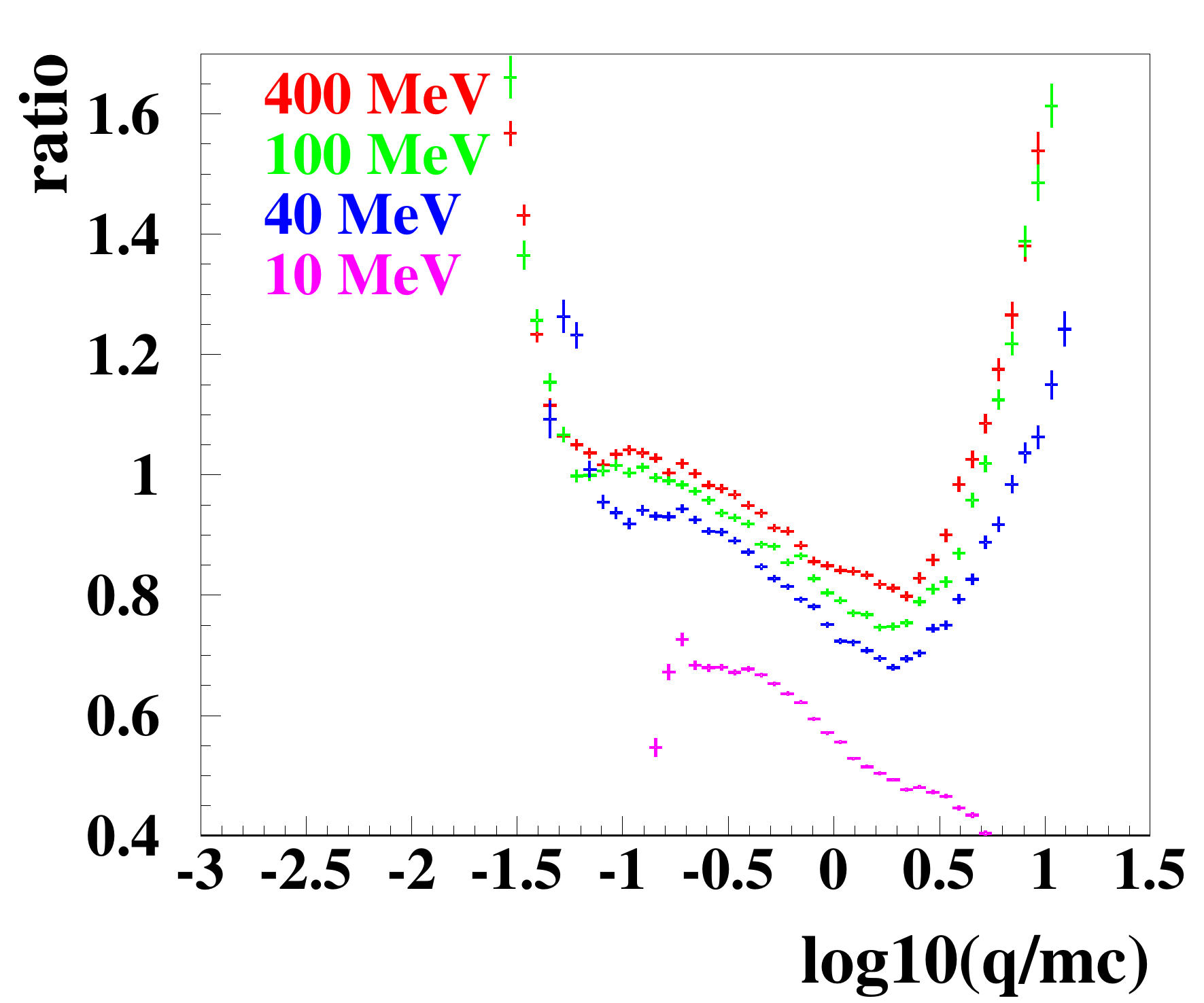}
}{
}
\caption{Ratio of the recoil momentum distribution 
 for triplet conversion to nuclear conversion. 
Without (left) and with (right) screening.
\label{fig:rapport:log10(q/mc)}
}
\end{figure}

The ratio of the distributions of the recoil momentum for triplet
conversion to nuclear conversion is shown in
Fig. \ref{fig:rapport:log10(q/mc)}.
When screening is not taken into account, a pattern similar to the
analytical computation by Mork \cite{Mork1967} is observed 
(Fig. \ref{fig:rapport:log10(q/mc)} left).
The difference between the effect of screening for triplet and nuclear
conversion is visible at low $q$ (Fig. \ref{fig:rapport:log10(q/mc)}
right).

\begin{figure}
\begin{center}
\iftoggle{couleur}{
 \includegraphics[width=0.8\linewidth]{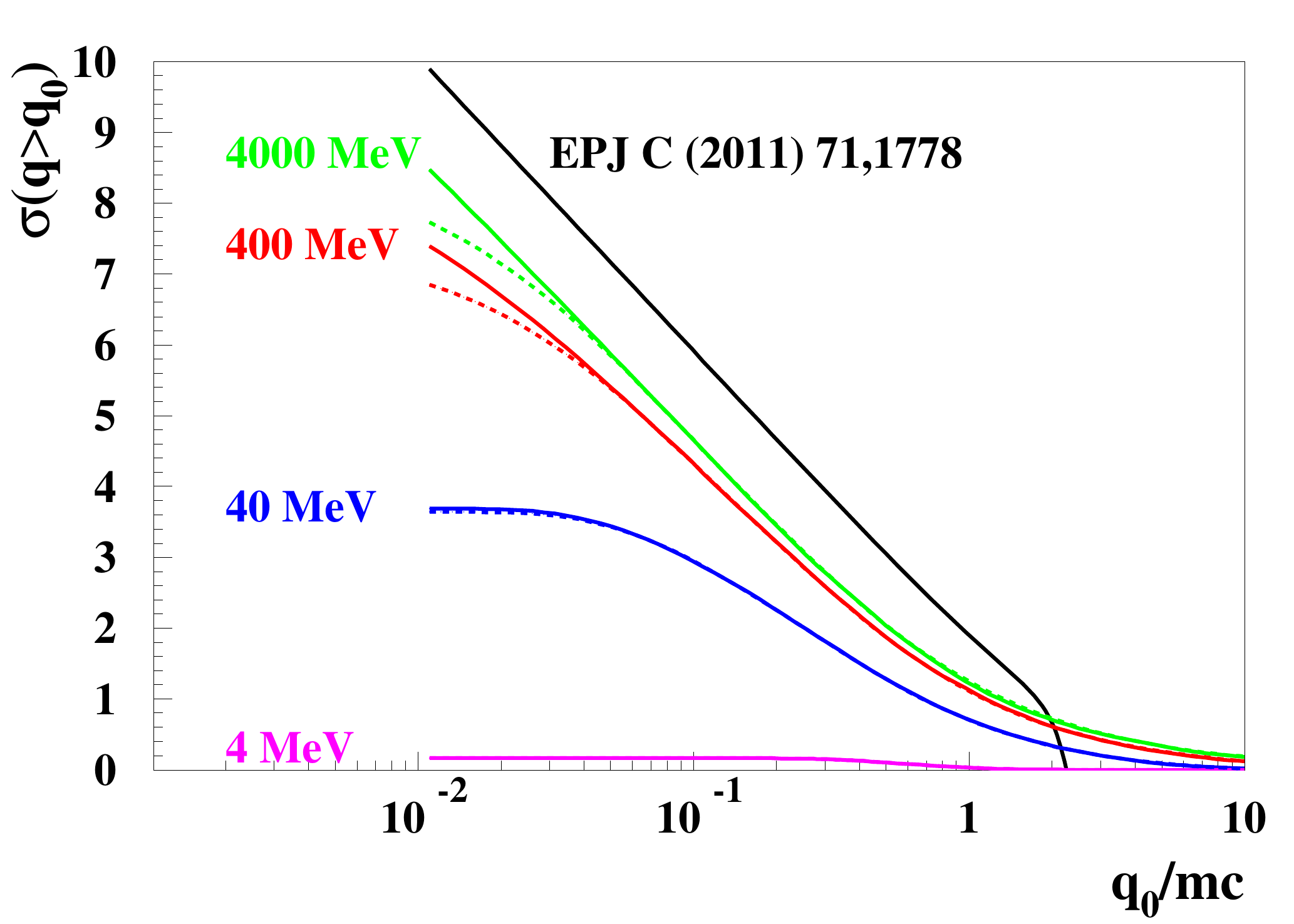}
}{
}
\caption{Variation of the triplet cross-section $\sigma(q>q_0)$ above
 a recoil momentum threshold, $q_0$, as a function of $q_0$ in
 units of $mc$, compared with the high-energy asymptotic expression of EPJC
 (2011) 71,1778 \cite{Iparraguirre:2011zz} (thick line).
With (dashed line) and without (solid line) form factor applied.
\label{fig:compare:Depaola}
}
\end{center}
\end{figure}

In a given experiment, the analysis of triplet conversion events can
be performed only for recoil momentum $q$ larger than some threshold $q_0$.
The magnitude of the cross-section for these events, $\sigma(q>q_0)$
is obviously a concern.
In the high-energy approximation, it was shown that $\sigma(q>q_0)$ is
energy independent \cite{Boldyshev:1994bs,Iparraguirre:2011zz}.
The way that the full calculation performed here tends to the high-energy
approximation \cite{Iparraguirre:2011zz} is shown in
Fig. \ref{fig:compare:Depaola}.

\section{Precision of the measurement of $P$}

When the measurement of the polarization is performed with a fit of
the $\phi$ distribution with the pdf of eq.~(\ref{eq:diff1D}), the
RMS resolution of the measurement of 
$\calA P$
is given by 
$\sigma_{(\calA P)} \approx \sqrt{2/N}$,
that is:
\begin{eqnarray}
\sigma_{P} \approx \frac{1}{\calA}\sqrt{\frac{2}{N}},
 \label{eq:P:sig}
\end{eqnarray} 
where $N$ is the number of events in the sample.

\begin{figure}
\begin{center}
\iftoggle{couleur}{
 \includegraphics[width=0.6\linewidth]{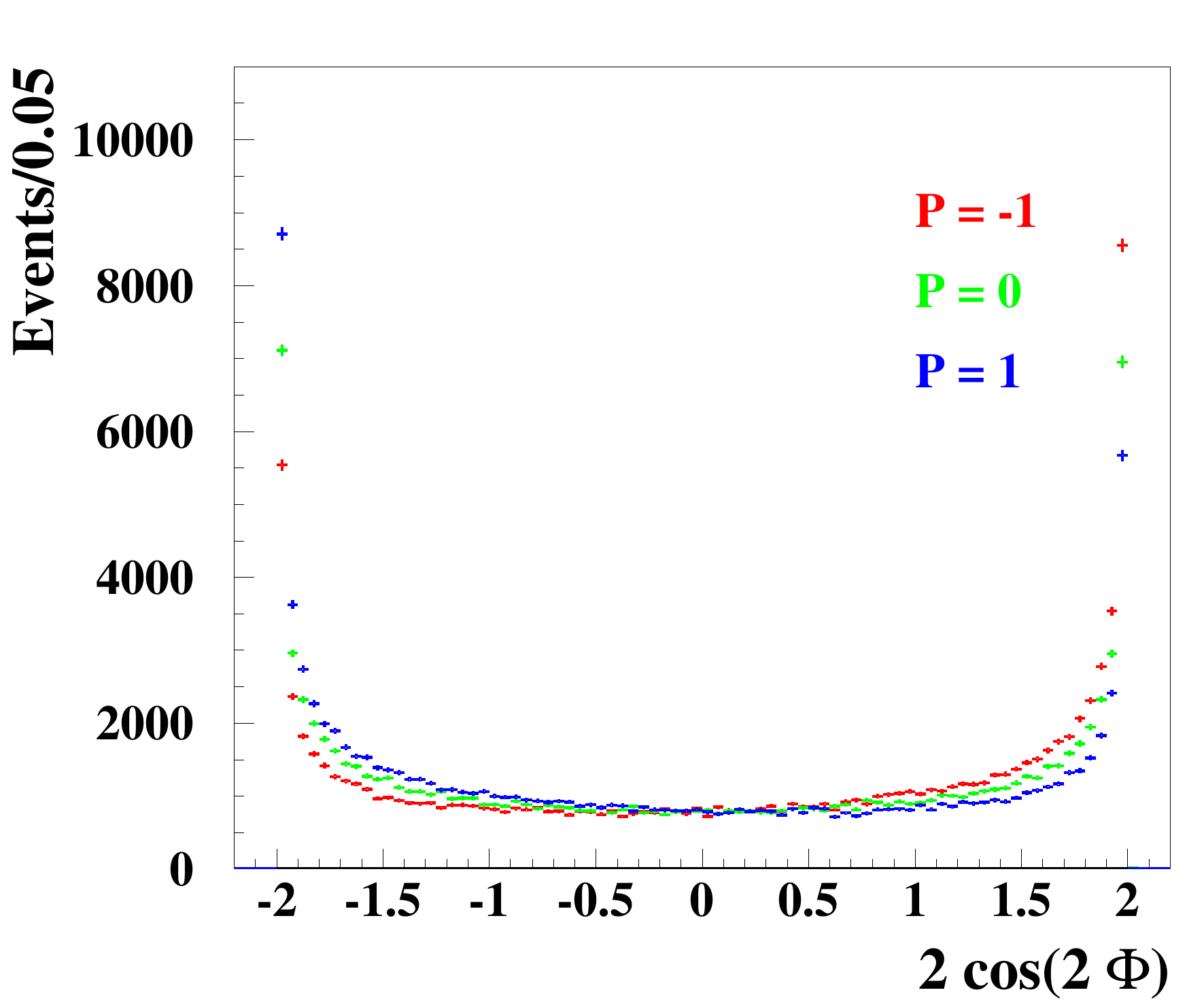}
}{
}
\caption{Distribution of $2 \cos(2 \phi)$ for the triplet conversion of 
40\,MeV photons.
\label{fig:2cos2phi}
}
\end{center}
\end{figure}

The value of $\calA P $ can also be computed from the moments of
appropriate weights\footnote{With this approach, the direction of the
 polarization of the source, that is the angle $\phi_0$
(see eq. \ref{eq:diff1D}), can be measured by a combined analysis of
weights $2\cos 2\phi$ and $2\sin 2\phi$.},
$w_{i}$, for event $i$, $i=1\cdots N$.
In general, the expectation value $\textE(w)$ of $w$ is a
function of the parameter(s) of the distribution (here of $P$).
We can then obtain an estimator of the parameter(s) from the measured
mean value
$\langle w \rangle$.
The expression for $\textE(w)$ reads: 
\begin{equation}
\textE(w) =\int \frac{w(\phi)}{\Gamma} \frac{\dd\Gamma}{\dd \phi} \dd \phi
\end{equation}
For $w=2\cos 2\phi$ (Fig. \ref{fig:2cos2phi}), we obtain 
$\textE(w) = \calA P$.
The uncertainty is obtained from the expression for the variance of
$w$
\begin{equation}
\sigma_{P} = \frac{1}{\calA\sqrt{N}} ~ \sigma_w, 
\label{eq:sigmaP}
\end{equation}

with the RMS $\sigma_w$ of $w$, 
$\sigma_w = \sqrt{\textE(w^{2}) - \textE(w)^{2}}$.
Here, $\textE(w^{2})=2$, so that 
\begin{equation}
\sigma_{P} = \frac{1}{\calA \sqrt{N}} \sqrt{2 - (\calA P)^{2}}, 
\end{equation}

\begin{figure}
\iftoggle{couleur}{
 \includegraphics[width=0.49\linewidth]{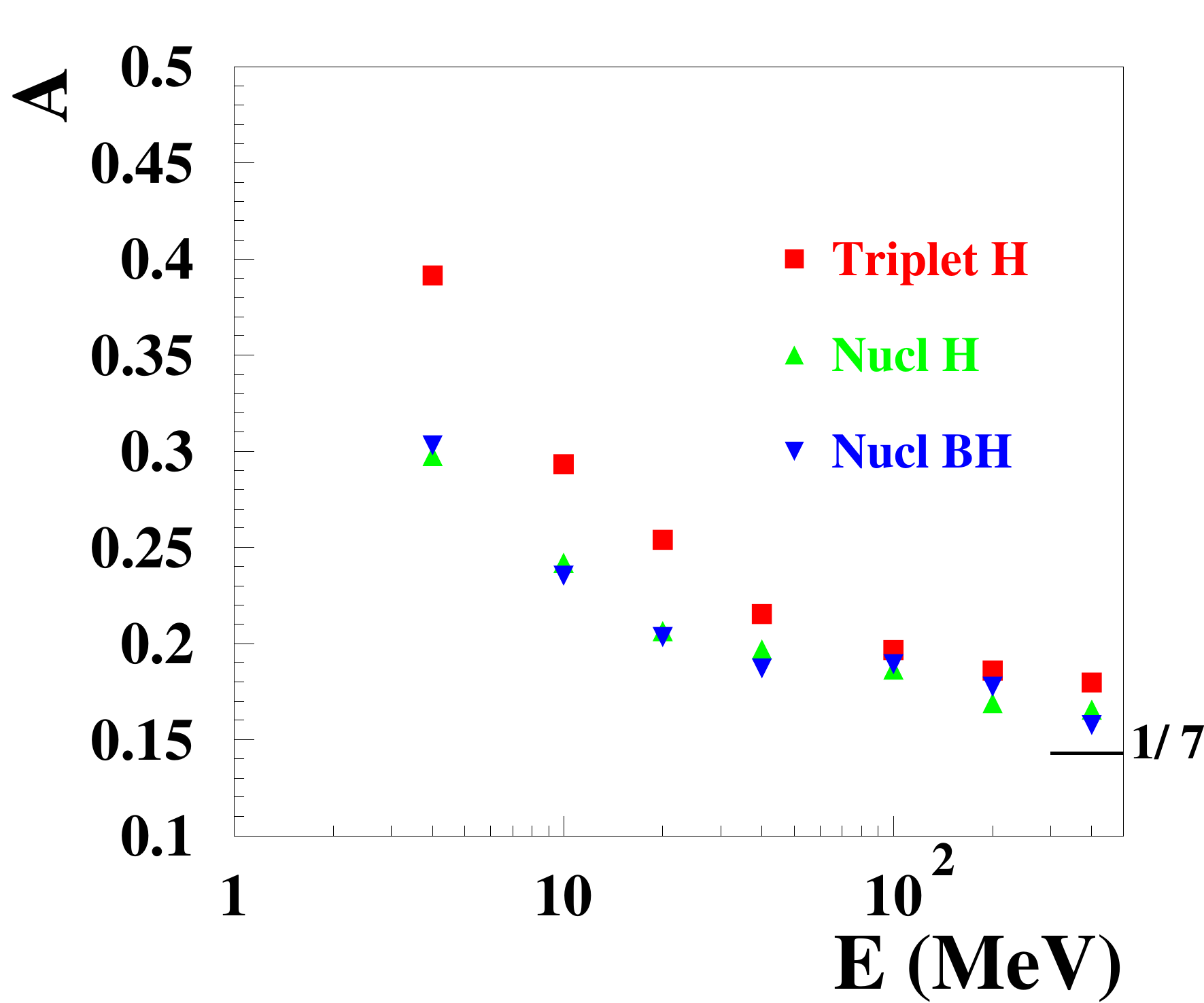}
 \includegraphics[width=0.49\linewidth]{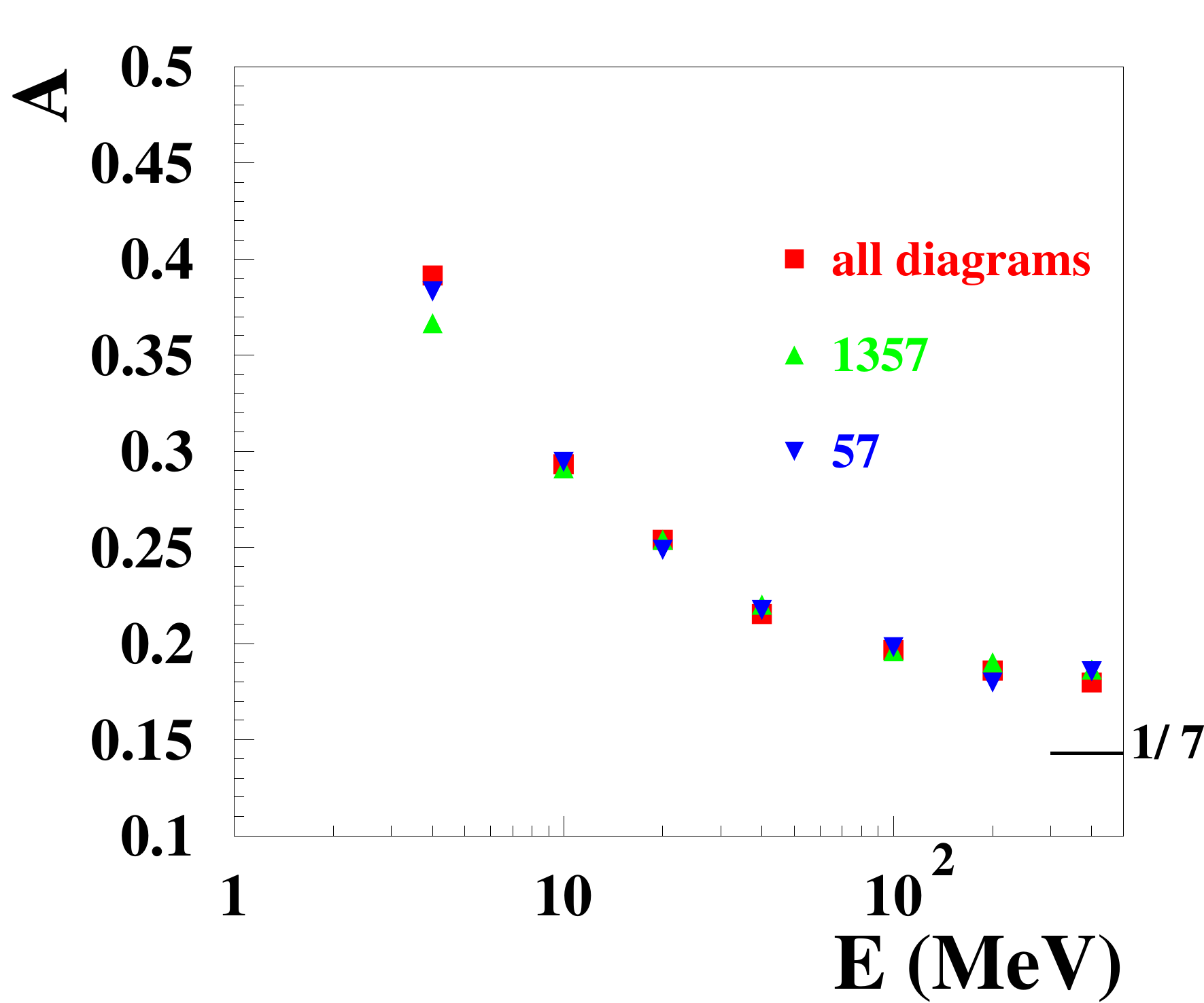}
}{
}
\caption{Left: Variation of \calA with photon energy
(no screening).
Right, for triplet conversion, all diagrams, without exchange diagrams (1357)
 and Bethe-Heitler only (57).
The high energy value from Ref. \cite{Boldyshev:1972va} is noted $1/7$.
\label{fig:lambda}
}
\end{figure}
\begin{figure}
\iftoggle{couleur}{
\centerline{\includegraphics[width=0.49\linewidth]{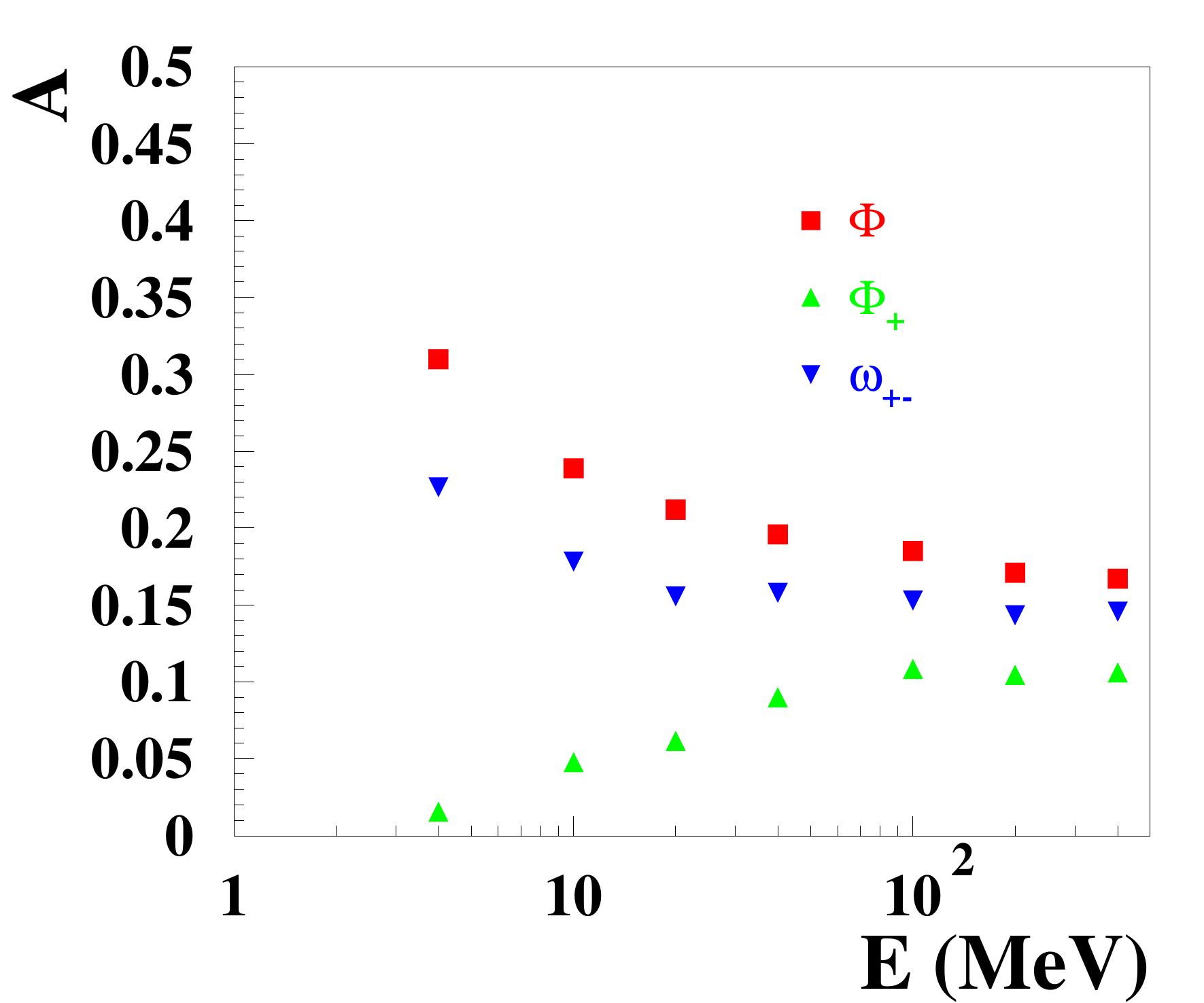}}
}{
}
\caption{
Variation of \calA , obtained from $\phi$, $\phi_+$ and
 $\omega_{+-}$, with photon energy for nuclear conversion
(no screening).
\label{fig:lambdaomega}
}
\end{figure}
\begin{figure}
\begin{center}
\iftoggle{couleur}{
 \includegraphics[width=\linewidth]{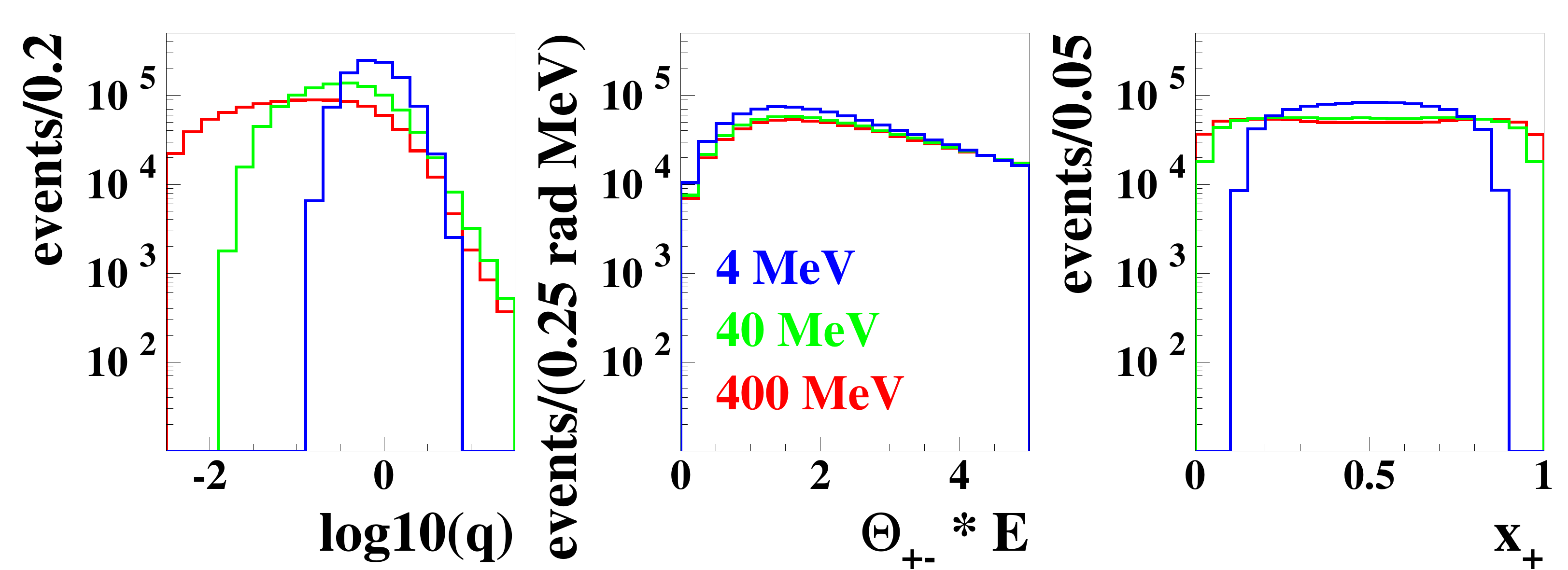}
\\
 \includegraphics[width=\linewidth]{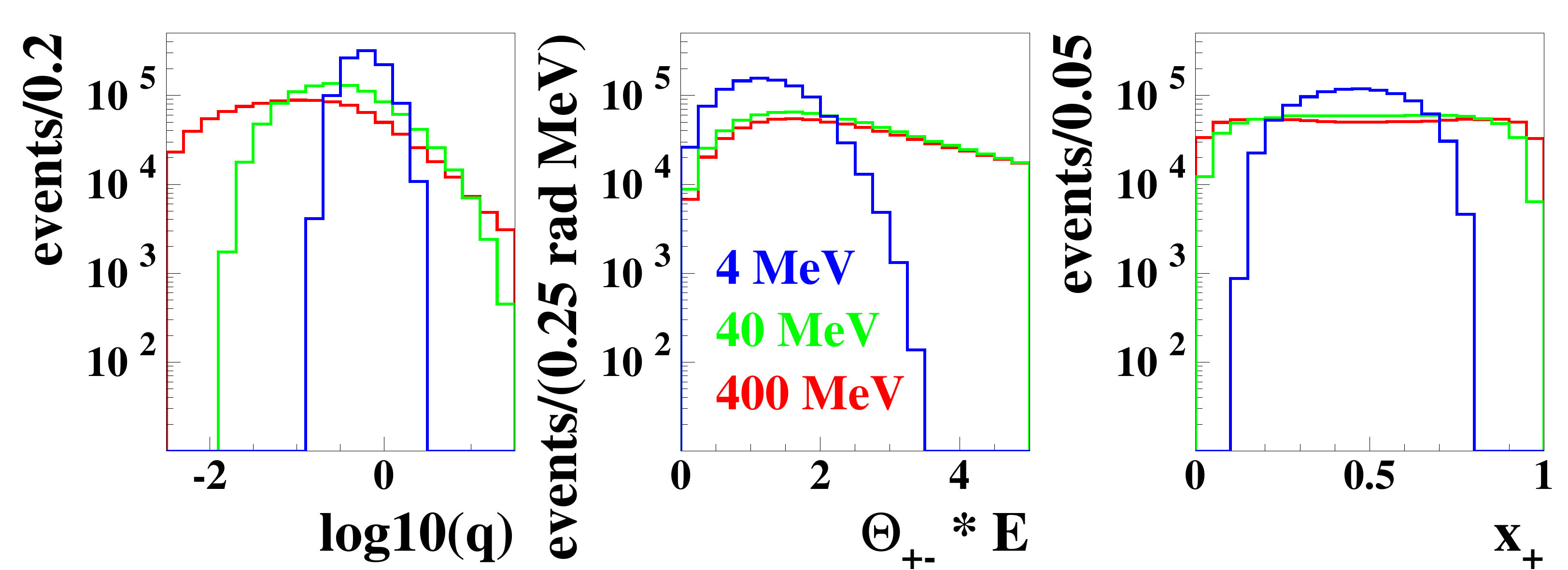}
}{
}
\caption{Spectra of 
 $\log_{10}(q(\mega\electronvolt/c))$, $\theta_{+-} \times E$ and $x_+$.
 Up: nuclear conversion, down: triplet conversion (no screening).
\label{fig:coupe-N}
}
\end{center}
\end{figure}


which is
smaller than the value obtained by the fit (eq.~(\ref{eq:P:sig})) and
that tends to it in the approximation of a small asymmetry and/or
polarization.
The variation of \calA obtained from $\textE( 2\cos(2 \phi))$ with
photon energy is shown in Fig. \ref{fig:lambda} (left).
In the above, the angle $\phi$ can be either the azimuthal angle of
the recoiling particle, or of the pair, that are back-to-back.
In the case of nuclear conversion though, since the recoil of the
nucleus goes undetected, $\phi$ cannot be measured
directly\footnote{Unless the source would be alone in the sky, or some
 extra information is used to determine the true direction of the
 incoming photon, such as time windowing for a GRB, or such as phase
 information for a pulsar.}.
The azimuthal angle of one of the tracks, e.g. that of the positron
$\phi_+$, may be used, in which case the effective polarization
asymmetry decreases \cite{Yadigaroglu:1996qq,Wojtsekhowski}. 
The decrease is extremely strong at low energy 
(Fig. \ref{fig:lambdaomega} right).
The angle that can be measured is the azimuthal angle that connects
the positron to the electron, $\omega_{+-}$
(Fig. \ref{fig:schema:angles}), as the direction of the incoming
photon is not known, in general.
The use of the angle $\omega_{+-}$ provides a partial recovery of the
sensitivity (\cite{Wojtsekhowski} and Fig. \ref{fig:lambdaomega} right).

In the case of triplet conversion, the three leptons of the final
state are reconstructed, and either angle can be used.

\subsection{Attempts to increase the polarization asymmetry}

There have been many attempts to make use of the variation of the
polarization asymmetry, \calA, with the kinetic variables that
describe an event, so as to increase the
effective value of \calA by a judicious event selection.
The asymmetry was found to be larger when the two electrons of the
pair share the energy equally\cite{Olsen:1959zz,Bakmaev:2007rs}, it
varies with the (azimuthal) acoplanarity, $\phi_+-\phi_-$, between the
two electrons of the pair\cite{Maximon:1962zz}, and it is larger for
small pair opening angles \cite{Endo:1992nq}.
This has lead to proposals of strategies that augment the mean value
of $\calA$ by an event selection.

\begin{figure}
\begin{center}
\iftoggle{couleur}{
 \includegraphics[width=\linewidth]{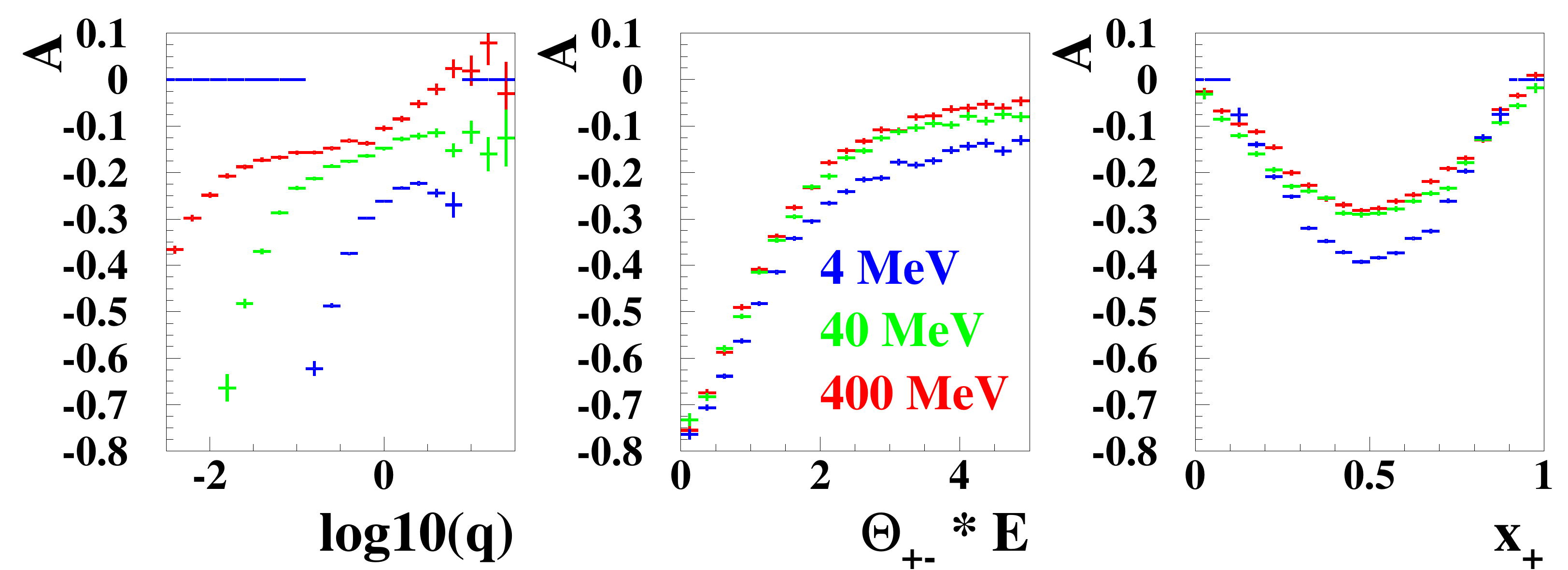}
\\
 \includegraphics[width=\linewidth]{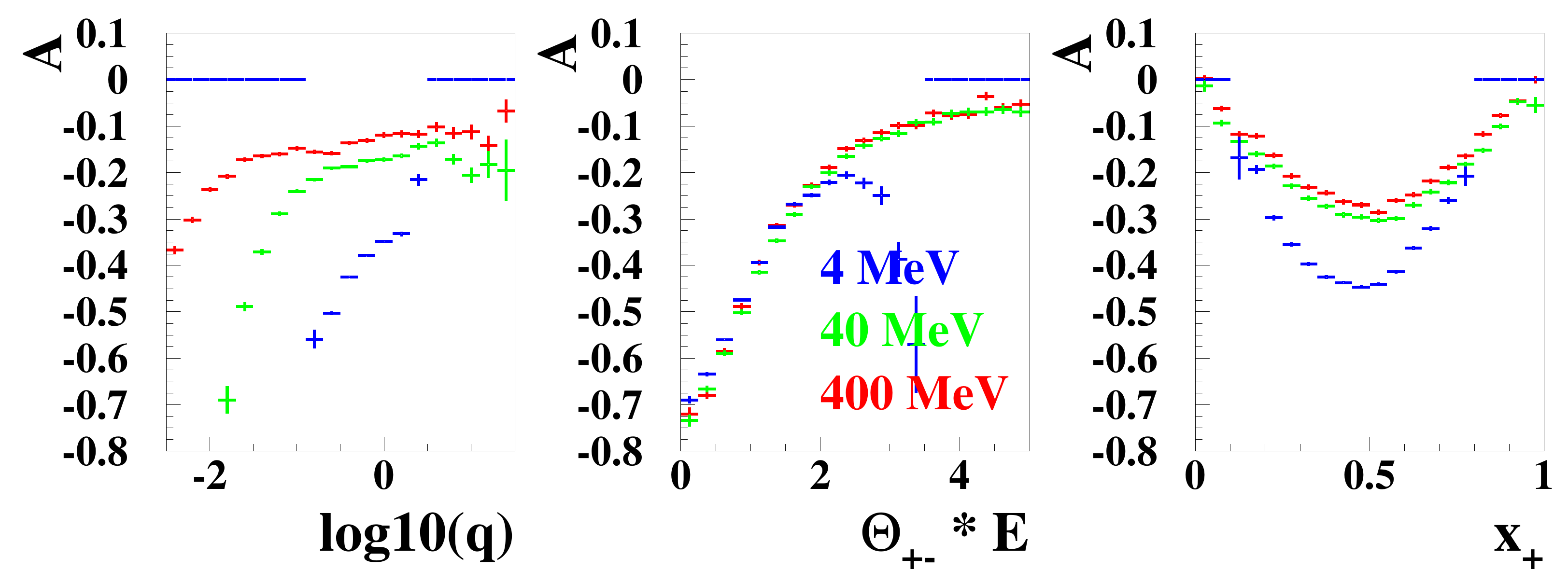}
}{
}
\caption{Variation of \calA measured as $\textE( 2\cos(2 \phi))$ 
(either the azimuthal angle of the pair or of the recoiling particle)
with $\log_{10}(q(\mega\electronvolt/c))$, $\theta_{+-} \times E$ and $x_+$.
 Up: nuclear conversion, down: triplet conversion (no screening).
\label{fig:coupe-A}
}
\end{center}
\end{figure}

Such a kinetic variable is generically noted $\chi$ in the following.
We find the polarization asymmetry \calA to be larger
at low (logarithm of the) recoil momentum $\log_{10}(q(\mega\electronvolt/c))$,
at low opening angle $\theta_{+-}$ (here rescaled by the photon
energy)
 and when the energy is balanced between the tracks of the pair
 ($x_{+} \approx 0.5$)
 (Fig. \ref{fig:coupe-A}).

\begin{figure}
\begin{center}
\iftoggle{couleur}{
 \includegraphics[width=\linewidth]{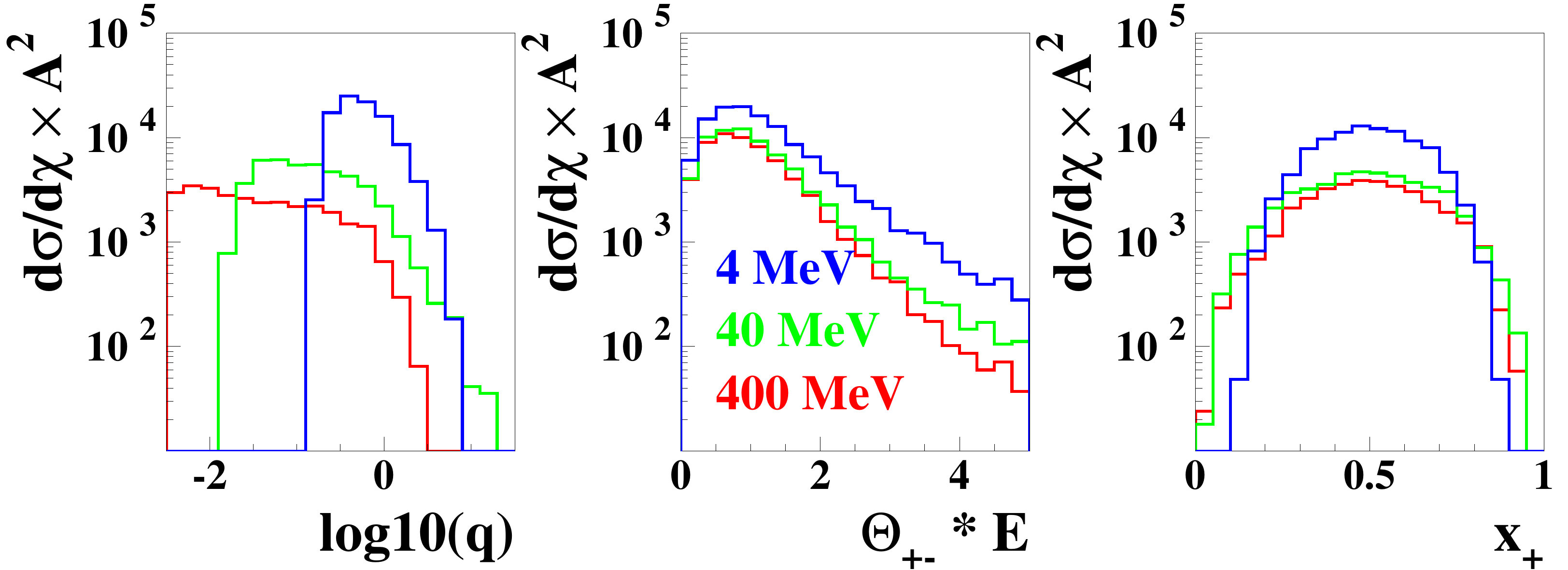}
\\
 \includegraphics[width=\linewidth]{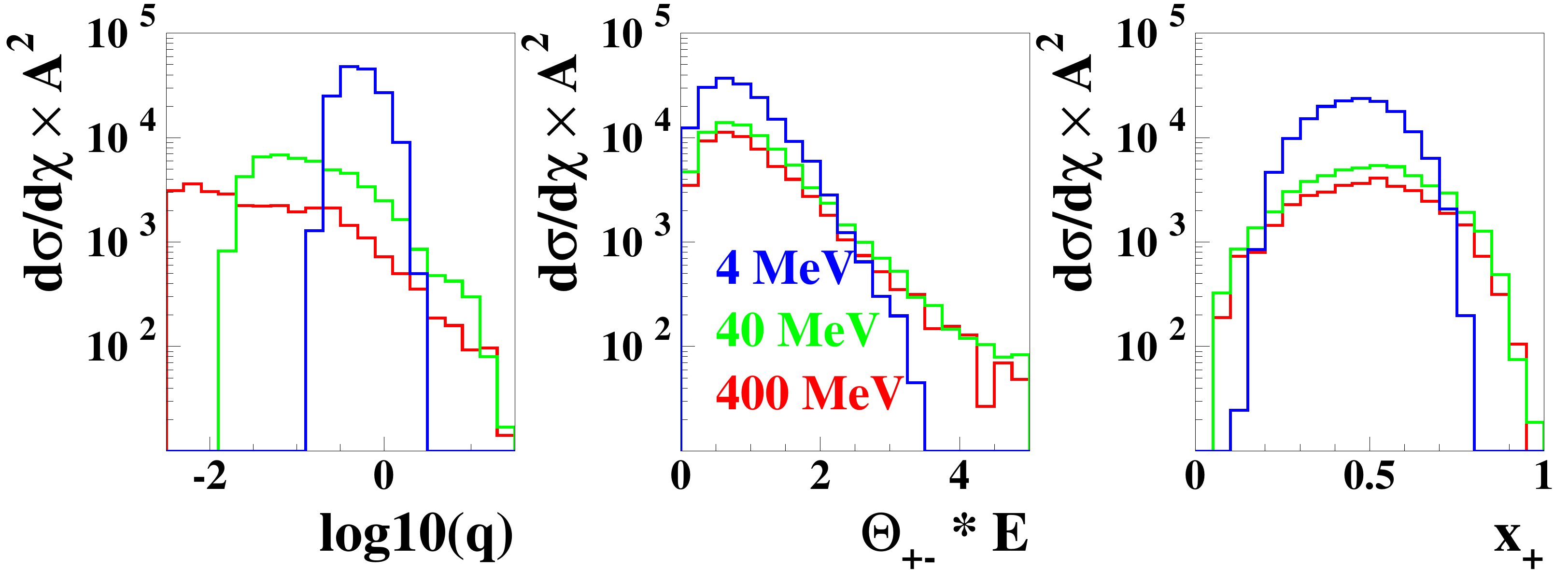}
}{
}
\caption{Variation of $F$
with $\log_{10}(q(\mega\electronvolt/c))$, $\theta_{+-} \times E$ and $x_+$.
 Up: nuclear conversion, down: triplet conversion (no screening).
\label{fig:coupe-F}
}
\end{center}
\end{figure}

Selecting events so as to increase the value of \calA for the selected
sample has a cost in terms of statistics and therefore in terms
of analyzing power.
The typical range of interest is where the figure of merit
$F \equiv \frac{\dd \sigma}{\dd \chi} \calA^2(\chi)$ 
is maximum (e.g. \cite{Endo:1992nq}\footnote{The numerical
 differences between the present results and those in
 Ref. \cite{Endo:1992nq} is due to their further selecting kinematic
 parameters according to a dedicated experiment -- in particular the
 opening angle peaks at a larger value of $7 m/E$ in
 Ref. \cite{Endo:1992nq}.}).
The variation of $F(\chi)$ for the three variables $q$, $\theta_{+-}$
and $x_+$ is shown in Fig. \ref{fig:coupe-F}.

We study such cuts here, making an event selection taking as an
example an upper side cut
 $\chi < \chi_c$ as is the case for variables $q$ and $\theta_{+-}$.
Optimizing such cuts involves minimizing the precision of the
measurement of $P$, eq.~(\ref{eq:sigmaP}), in which the factors
\calA , $N$ and $\sigma_w$ are estimated with the cut applied.
In practice we use here its inverse, normalized to an initial total
sample of one event, and we note this new figure of merit $G$:
\begin{equation}
G(\chi_c)
=
\frac{\calA(\chi_c)}{\sigma_w(\chi_c)}
\sqrt{\frac{N(\chi_c)}{N}}.
\label{eq:G}
\end{equation}

\begin{figure}
\begin{center}
\iftoggle{couleur}{
 \includegraphics[width=\linewidth]{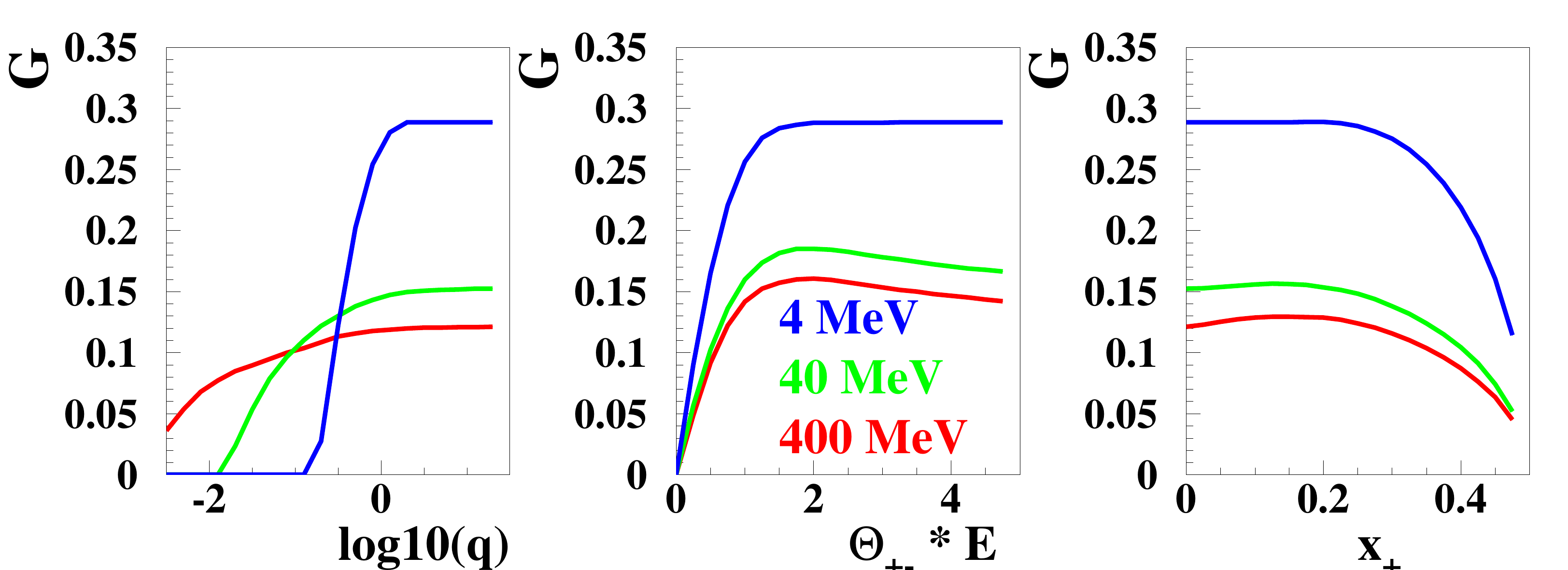}
}{
}
\caption{Variation of $G$ with cut position of variables 
 $\log_{10}(q(\mega\electronvolt/c))$, $\theta_{+-} \times E$ and $x_+$
 for triplet conversion (no screening).
\label{fig:G}
}
\end{center}
\end{figure}

The variation of the figure of merit $G$ with $\chi_c$
(Fig. \ref{fig:G}) shows that this selection strategy brings very
little, if any, improvement.
For example, for triplet conversion at photon energy of 40\,MeV, a
$\theta_{+-}\times E < 2\,\radian \cdot\mega\electronvolt$ cut would result
in an increase of the effective asymmetry from 21.3 to 40.6 \% at the
cost of a cut efficiency of 38\%, with a small $G$ improvement from 0.153 to
0.185.

\subsection{The use of an optimal variable}

A way to improve the polarization sensitivity, beyond the simple cuts
used above, is to make an optimal use of the information contained in
the multi-dimensional probability density function (pdf) of the
signal, through the use of an optimal variable
(see e.g.
\cite{Atwood:1991ka,Davier:1992nw,HenrotVersille:2000xr,Tkachov:2000xq}).
We note $\Omega$, the set of kinematic variables that fully describes
an event, $p(\Omega)$ its pdf, that depends on the parameter (here
$P$) that we want to measure.
We are looking for a weight $w(\Omega)$, 
so that the $P$ dependence of $\textE(w)$ allows a measurement of $P$, and
that the variance of such a measurement is minimal.
The solution is 
\begin{equation}
w_{\textroman{opt}} = \frac{\partial \ln p(\Omega)}{\partial P}.
\end{equation}

In the particular case of a polarization measurement 
\begin{equation}
p(\Omega) \equiv f(\Omega) + P \times g(\Omega), 
\label{eq:p(omega)f+Pg}
\end{equation}
with $\int f(\Omega) \dd \Omega = 1$ and $\int g(\Omega) \dd \Omega = 0$,
we obtain 
\begin{equation}
w_{\textroman{opt}} = \frac{g(\Omega)}{f(\Omega) + P \times g(\Omega)}.
\label{eq:w:opt}
\end{equation}

In this study, we rescale the weight by a factor of 2, so
as to provide an asymmetry that is commensurate with \calA .
Also, 
in the approximation that the polarization is small
\cite{Atwood:1991ka}, 
we can neglect the $ P \times g(\Omega)$ term in the denominator, and obtain 
\begin{equation}
w_0 = 2 \frac{g(\Omega)}{f(\Omega)}.
\label{eq:w:0}
\end{equation}

In the case of the reduced equation (eq.~(\ref{eq:diff1D})) we find:
\begin{equation}
w_1 = 2 \cos{2\phi}, 
\end{equation}
a weight that therefore makes optimal use of the reduced information
present in the 1D distribution.

\begin{figure}
\begin{center}
\iftoggle{couleur}{
 \includegraphics[width=\linewidth]{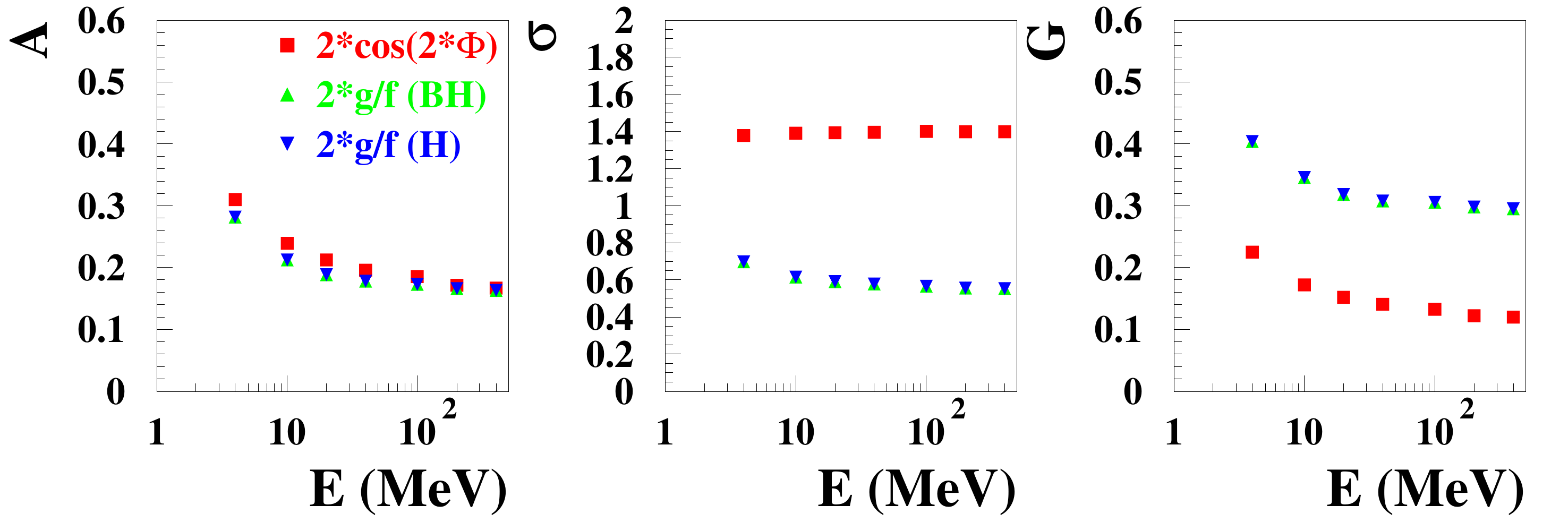}
\\
 \includegraphics[width=\linewidth]{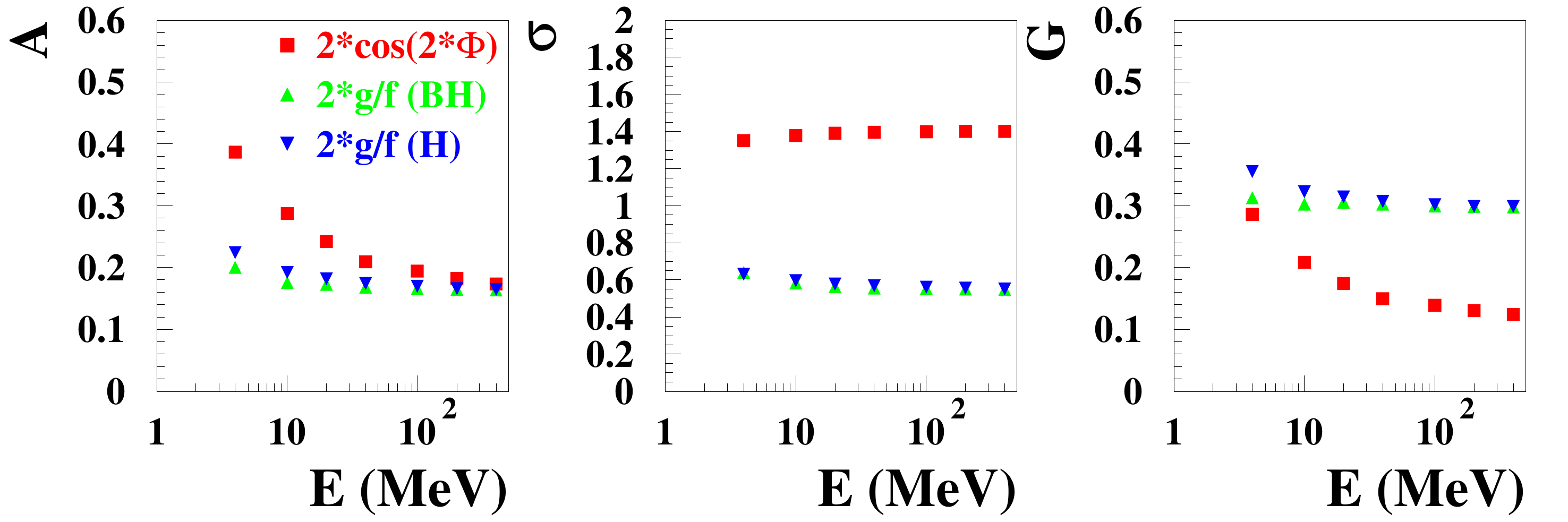}
}{
}
\caption{Comparison of the performances of 1D and 5D weights on the same sample.
 Up: nuclear conversion, down: triplet conversion (no screening; plots with screening are very similar).
\label{fig:1d5d}
}
\end{center}
\end{figure}

We can now estimate the loss of information when using the 1D reduced
distribution by comparing the performance of the 1D weight $w_1$ with that of
the 5D weights $w_0$.
In the case of the Bethe-Heitler (BH) pdf, $f(\Omega)$ and
$g(\Omega)$ are simply obtained from eq.~(\ref{eq:p(omega)f+Pg}).
In the case of the HELAS computation of the pdf (H), the unpolarized
part $f(\Omega)$ is obtained as the average of an $x$-polarized and
$y$-polarized differential cross-section.
Both calculations of the nuclear 5D weights yield results that
agree
perfectly with each other (Fig. \ref{fig:1d5d}), while a slight
discrepancy between BH and H is visible at low energy for triplet
conversion.
We can see that at a given photon energy, the three weights provide
similar values of the asymmetry, while the uncertainty is improved by
a factor larger than 2 for a 5D optimal weight, resulting in an
increase of $G$ by the same factor.
If it were
 possible to measure the kinematic variables that describe
the final state of an event with sufficient precision, using a 5D
optimal weight would therefore improve the precision of the
measurement of $P$ by up to a factor of about 2.

\begin{figure}
\begin{center}
\iftoggle{couleur}{
 \includegraphics[width=\linewidth]{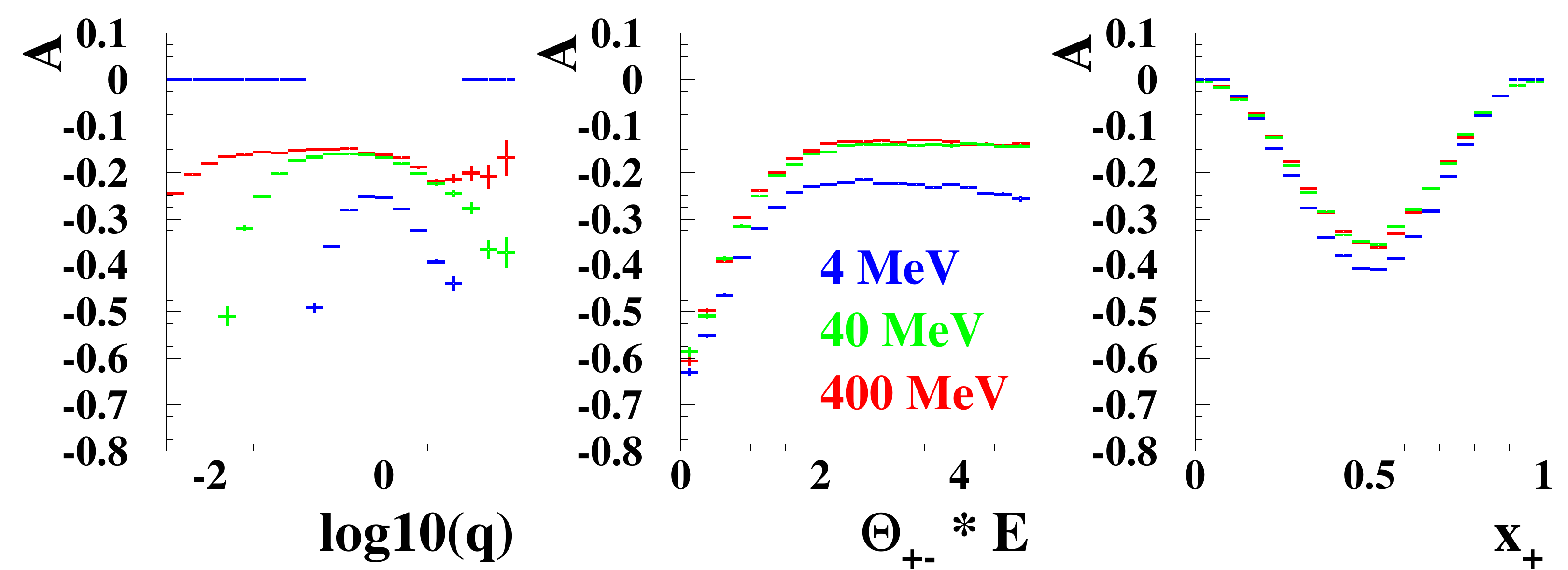}
\\
 \includegraphics[width=\linewidth]{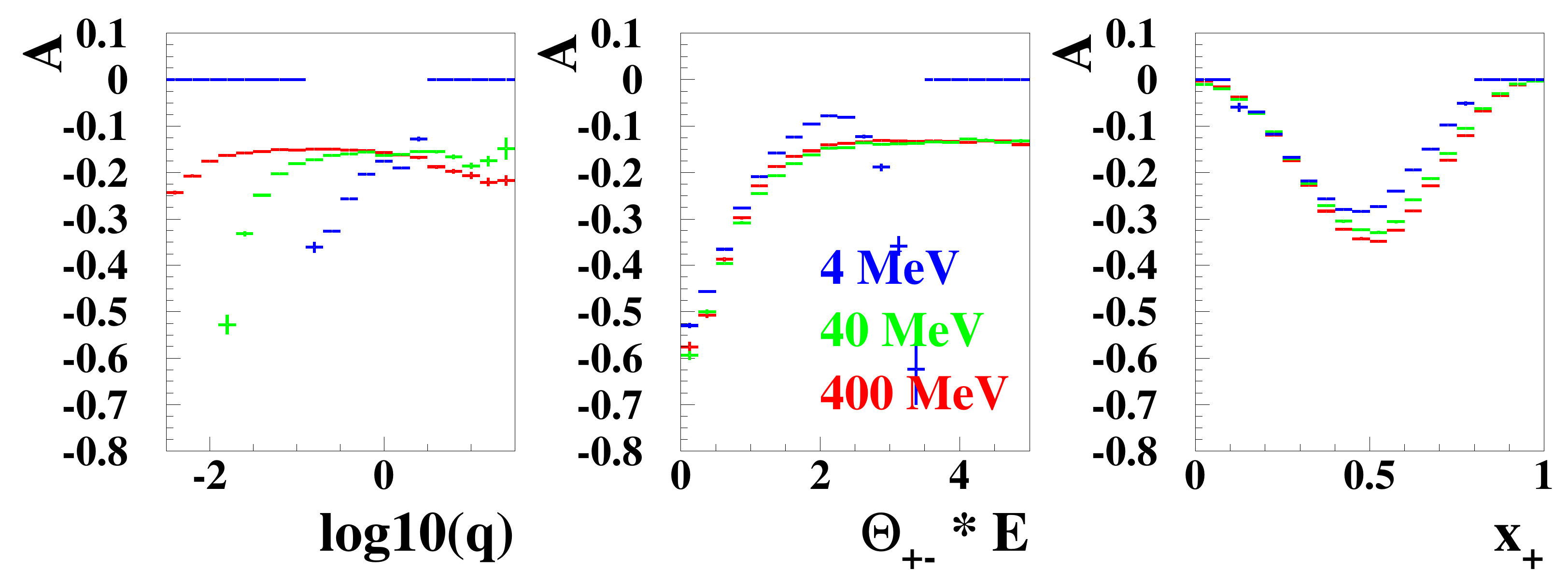}
}{
}
\caption{Variation of \calA measured as $\textE(w_0)$ 
with $\log_{10}(q(\mega\electronvolt/c))$, $\theta_{+-} \times E$ and $x_+$.
 Up: nuclear conversion, down: triplet conversion (no screening).
\label{fig:hop-coupe-A}
}
\end{center}
\end{figure}

We
note that $\textE(w_0)$ goes to an asymptote at high $q$ and
$\theta_{+-}$ (Fig. \ref{fig:hop-coupe-A}), in contrast with
$\textE(w_1)$ for which it keeps on decreasing to zero
(Fig. \ref{fig:coupe-A}).

After having characterized various ways to measure $P$ at generator
level, we now consider the consequences of the experimental effects on
the measurement of $P$.

\section{Experimental effects}
\label{sec:expt:effects}

\subsection{Nuclear conversion}

We first re-examine the configuration that has been studied in the
literature, i.e. a detector composed of a series of converter slabs
in which photons convert, interleaved with tracking detectors in
which electrons are tracked.
We then turn to an active target, i.e. a single homogeneous detector
 which at the same time converts photons and tracks electrons.
For this study of experimental effects in nuclear conversion, the 1D
weight is computed with angle $\omega_{+-}$, which is the quantity
that is available in practice.

\subsubsection{Slab converter}

For the first case (a slab detector), tracking is performed after the
tracks exit the converter.
The angular resolution for these tracks is therefore at least equal to
the average deflection angle over their path length $x$ inside the
slab\footnote{We consider here that the tracks traverse the same full
 thickness $x$, i.e. we don't consider the fluctuation of the
 conversion point.}.
The order of magnitude of the effect on the measurement of $\phi$ is
first estimated following Refs. \cite{Kotov,Mattox}, updating the
numerical values they used to parametrize the multiple
scattering.
In the small angle approximation, the $\phi$ angle resolution can be
expressed as
\begin{equation}
\sigma_{\phi} = \gfrac{\theta_{0,e^+} \oplus \theta_{0,e^-}}{\theta_{+-}}, 
\label{eq:sigmaPhi-0-slab}
\end{equation}
where $\theta_{0,e^+}$ and $ \theta_{0,e^-}$ are the 
RMS width of the projected deflection distribution of the positron and
of the electron, respectively, and $\theta_{+-}$
is the opening angle of the pair.
$\theta_{0}$ is obtained from the Gaussian approximation of
multiple scattering \cite{Beringer:1900zz}: 
\begin{equation}
\theta_{0}=
 \gfrac{p_0}{\beta p} \sqrt{\gfrac{x}{X_0}},
\label{eq:theta:MS:PDG}
\end{equation}
where $p_0 = 13.6\,\mega\electronvolt/c$ and the small logarithmic
correcting factor has been neglected.

\paragraph{Analytical approximate expression, using the most probable opening angle} 

We first use the approximation made in Refs. \cite{Kotov,Mattox} of an
opening angle $\theta_{+-}$ equal to its most probable value
$\hat{\theta}_{+-} \approx E_0/E$.
At equipartition, $p_- \approx p_+$, we obtain:
\begin{equation}
\sigma_{\phi} 
\approx 
\gfrac{2 \sqrt{2} ~ p_0}{E_0} \sqrt{\gfrac{x}{X_0}}
\approx \sigma_0 \sqrt{\gfrac{x}{X_0}}, 
\label{eq:sigmaPhi-1-slab}
\end{equation}
where $ \sigma_0 \equiv 2 \sqrt{2} p_0 / E_0 \approx 24\,\radian$.
We note that the expression for $\sigma_{\phi}$ is independent of $E$:
at higher energy the track suffers less multiple scattering but the
pair opening angle gets smaller.
Taking the small logarithmic correcting factor into account in the
expression for $\theta_{0}$ for that particular value of the thickness
leads to a 28\% decrease of $\sigma_0$ (to about 17 \radian) and of
$\sigma_{\phi}$.
The critical thickness $x_c$ that induces a dilution of the polarization
asymmetry of a factor of 2 is therefore such that 
the normalized thickness
$t_c \equiv x_c / X_0 = ( \sigma_c / \sigma_0 )^2 \approx 1.2 \times 10^{-3} $,
which amounts, for example, to $110\,\micro\meter$ of silicon.

\paragraph{Simulation} 

We now use the full 5D simulation.
\begin{figure*}
\begin{center}
\iftoggle{couleur}{
 \includegraphics[width=\linewidth]{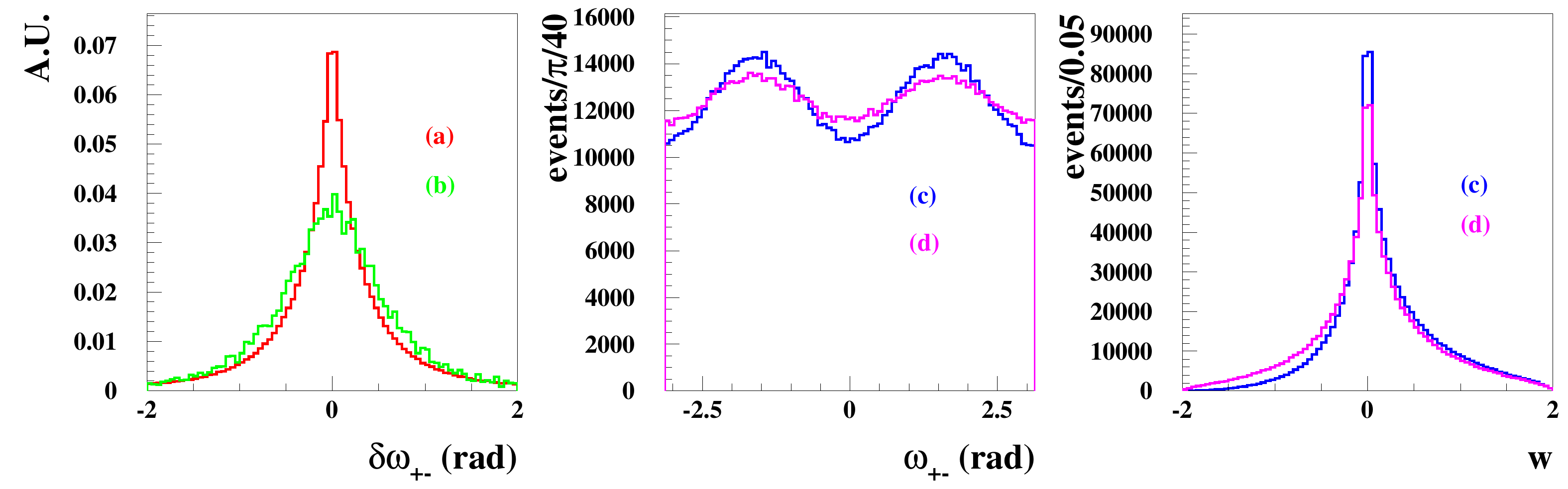}
}{
}
\caption{Left: distribution of the azimuthal angle resolution
$\delta \omega_{+-}$ due to multiple scattering through a 
$t_c = 1.7 \times 10^{-3} $ normalized thickness slab converter
 (nuclear conversion of 10\,MeV photons).
(a): bare generator with $|\delta \omega_{+-}|< 2\,\radian$;
(b): event selection as 
$|\theta_{+-}\times E - E_0 |< 0.2\,\mega\electronvolt$, 
 $0.8 < E_+/E-< 1.2$, 
 $|\delta \omega_{+-}|< 2\,\radian$.
Center: 
Azimuthal angle $\omega_{+-}$ distribution; (c) bare generator;
(d) with multiple scattering.
Right: 
distribution of weight $w_0$ as defined in eq.~(\ref{eq:w:0})
 with the BH differential cross-section.
\label{fig:omega-residual}
}
\end{center}
\end{figure*}
We first cross-validate the above analytical results and the
simulation, parametrizing the multiple scattering with
the approximation of 
eq.~(\ref{eq:theta:MS:PDG}), and 
selecting events with a pair opening angle close to the maximum,
$|\theta_{+-}\times E - E_0 |< 0.2\,\mega\electronvolt$, 
and energy sharing close to equipartition, $0.8 < E_+/E-< 1.2$: 
we obtain an RMS width for $\omega_{+-}$ of $0.581 \pm 0.005\,\radian$, 
which is compatible with that of $\sigma_c $.
Releasing the selection produces a more peaked $\omega_{+-}$ residual
distribution, with a similar RMS, but that is smaller by about 8\%.
Using the full expression for the multiple scattering 
 \cite{Beringer:1900zz} further decreases the RMS. 
\begin{figure}
\begin{center}
\iftoggle{couleur}{
 \includegraphics[width=0.6\linewidth]{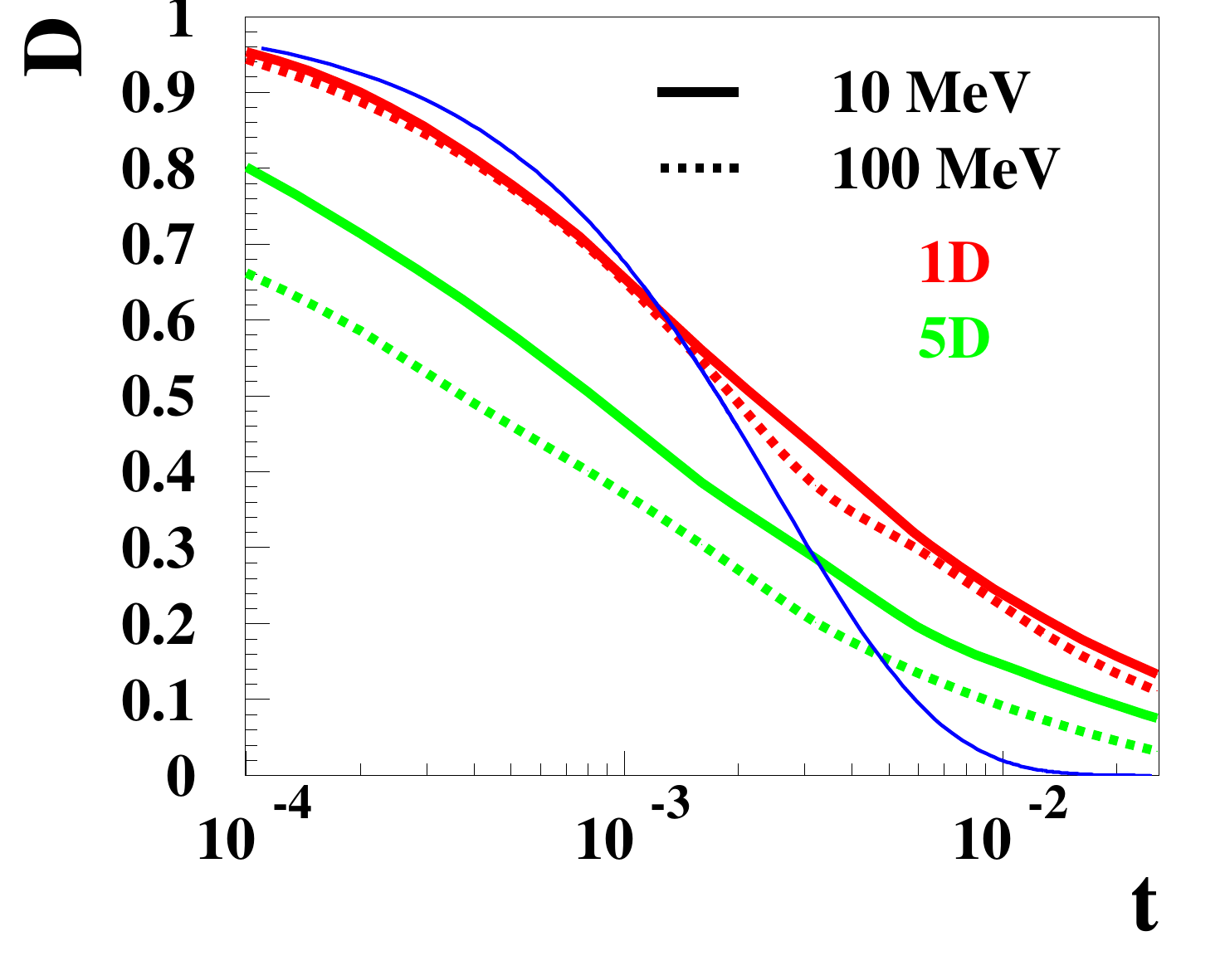}
}{
}
\caption{Variation of the dilution factor $D$ with the slab normalized
 thickness $t$, the polarized fraction $P$ being estimated with the
 expectation value of either the 1D weight or of the 5D weight,
 for photon energies of 10 and 100\,MeV (nuclear conversion).
The thin line shows the approximation using the most probable opening
angle ($\sigma_0 = 14\,\radian$).
\label{fig:dilution}
}
\end{center}
\end{figure}
Finally an RMS equal to $\sigma_c $ is obtained for a normalized
thickness of $t_c \approx 1.7 \times 10^{-3} $,
that is for $\sigma_0 = \sigma_c /\sqrt{t_c} \approx 14.6\,\radian$.
The dilution of the polarization asymmetry so induced is visible in
Fig. \ref{fig:omega-residual} center and right.
Since
the distribution of the azimuthal angle residual is non
Gaussian in the full simulation (Fig. \ref{fig:omega-residual} left),
the dilution factor can be different from 1/2 for an RMS equal to
$\sigma_c$.

The variation of the dilution factor, $D$, with the normalized
thickness, $t$, is shown in Fig. \ref{fig:dilution}.
The dilution factor is smaller for the 5D weight than for the 1D
weight, which indicates that the track angular resolution affects not
only the azimuthal angle but also the detail of the 5D differential
cross-section.
Note that in the case of the 5D weight, the dilution is also affected by
effects that are not taken into account in the present study, such as
the resolution of the measurement of the energy-momentum
of each track. 
Also the single-track angular resolution and the non-observation of
the recoil in the case of nuclear conversion induce an uncertainty in
the definition of the $z$ axis.
The simpler estimator based on the weight computed from angle
$\omega_{+-}$ is immune from these effects, and therefore we will use
the 1D weight as a conservative benchmark of the precision on $P$.

The variation of $D(t)$ obtained with the simulation, which includes
the full $\theta_{+-}$ distribution, turns out to be very different
from the expression $D(t) = e^{-2\sigma_0^2 \times t}$ based on the particular value
$\hat{\theta}_{+-}$ of $\theta_{+-}$.

One might be tempted to select events with large opening angle
$\theta_{+-}$ in the hope of getting a better dilution, but 
for $\textE(w_0)$, the asymmetry is not improved, and we merely get
a loss in statistics.
For $\textE(w_1)$ it is even worse (plots not shown), as the
asymmetry itself decreases at high $\theta_{+-}$
(Fig. \ref{fig:coupe-A}).

\subsubsection{Thin detectors}

\begin{figure}
\begin{center}
 \includegraphics[width=0.8\linewidth]{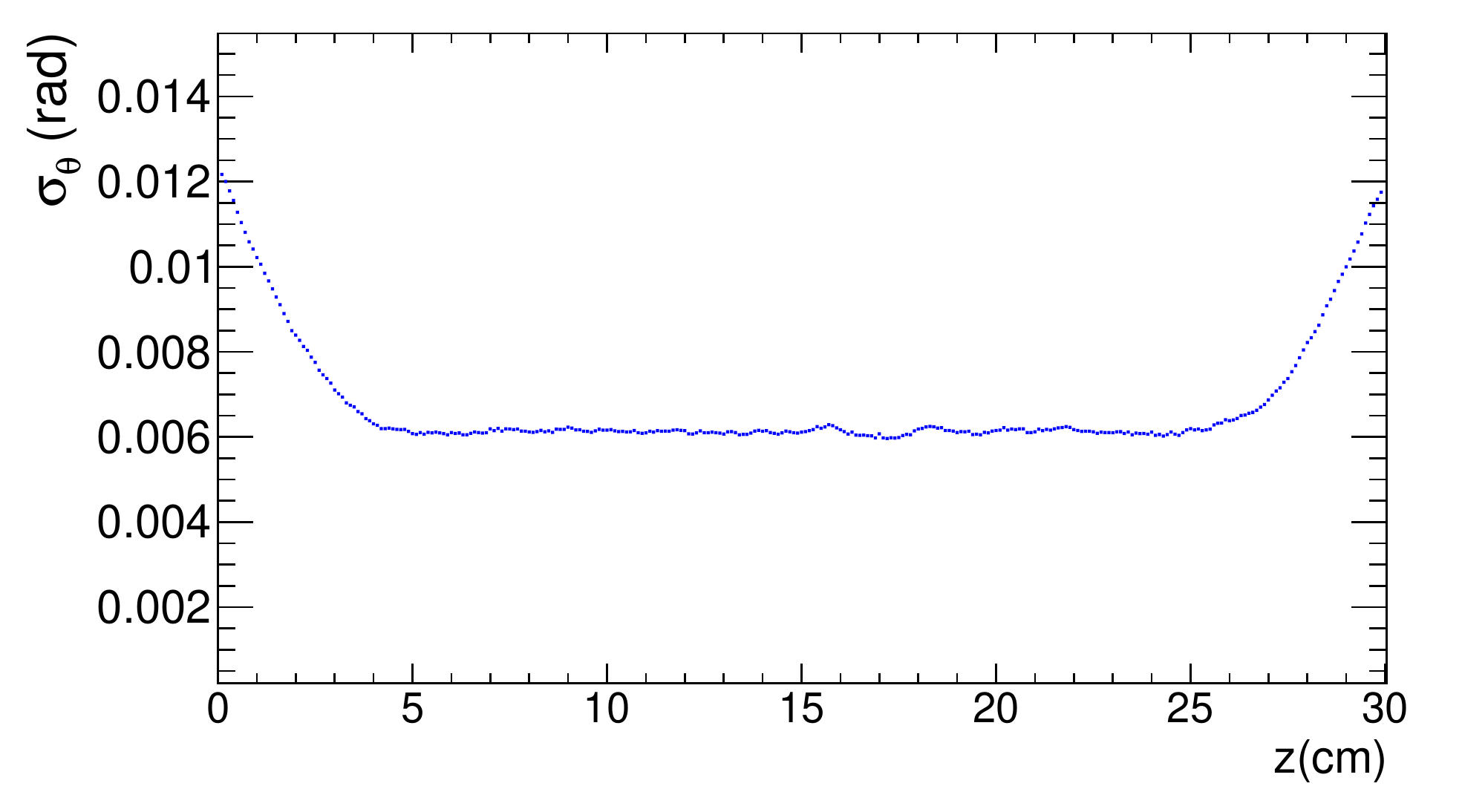}
\caption{Single track angle RMS uncertainty of a Kalman filter fit
 from a sample of 5000 simulated $40\,\mega\electronvolt/c$ tracks in
 5\,bar argon with $\sigma = l = 0.1 \centi\meter$;
Eqs. (\ref{eq:p_1}) predicts an RMS of $12.2\,\milli\radian$
(From Ref. \cite{ShaoboWang}).
\label{fig:KalmanValidation}
}
\end{center}
\end{figure}

In a thin detector, the optimal, minimal value of the
single-track angular resolution in the low-momentum,
multiple-scattering-dominated regime is given by
\cite{Bernard:2012uf}:
\begin{equation}
\sigma_{\theta tL} 
 \approx 
(2 \sigma)^{1/4} l^{1/8} X_0^{-3/8} (p/p_0)^{-3/4}
=
(p/p_1)^{-3/4},
 \label{eq:Vbb:p:2}
\end{equation}

with:
\begin{equation}
p_1 = p_0 \left( \gfrac{4 \sigma^2 l}{X_0^3} \right)^{1/6},
 \label{eq:p_1}
\end{equation}

where $l$ is the size of the longitudinal sampling of the detector,
and $\sigma$ is its single point spatial resolution.
Typical values for $p_1$ range from $\kilo\electronvolt/c$ to 
 $\mega\electronvolt/c$ with $p_1 = 50\,\kilo\electronvolt/c$
for 1\,bar argon, $p_1 = 1.45\,\mega\electronvolt/c$ for liquid argon
($\sigma = l = 0.1 \centi\meter$).
Equations (\ref{eq:Vbb:p:2}), (\ref{eq:p_1}) were obtained
\cite{Bernard:2012uf} from the expressions for optimal fits in the presence
of multiple scattering \cite{Innes:1992ge}.
Such fits can be implemented by using a Kalman filter
(Fig. \ref{fig:KalmanValidation}, \cite{Fruhwirth:1987fm}).

\paragraph{Analytical approximate expression, using the most probable opening angle} 

\begin{figure}
\begin{center}
\iftoggle{couleur}{
 \includegraphics[width=0.8\linewidth]{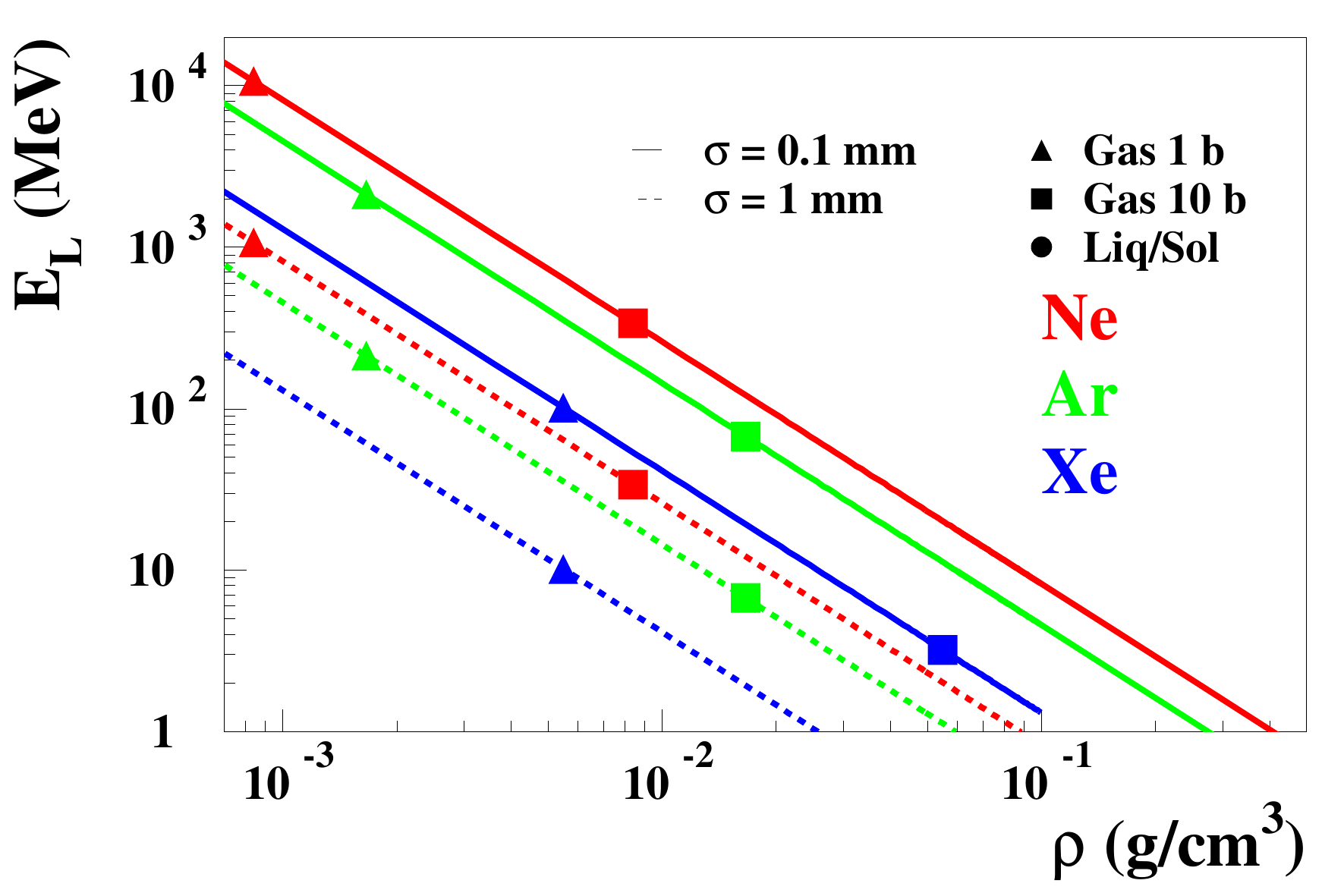}
}{
}
\caption{Variation of the critical energy, $E_L$, up to which
 polarimetry in a TPC is possible, with TPC density (thin detector,
 nuclear conversion).
Bullets that correspond to liquid / solid TPC are not visible as the value of $E_L$ is smaller than 1 MeV.
\label{fig:E_L}
}
\end{center}
\end{figure}

Again, we first consider the most probable value $\hat{\theta}_{+-}$
of $\theta_{+-}$.
From eqs. (\ref{eq:sigmaPhi-0-slab}), (\ref{eq:Vbb:p:2}) and (\ref{eq:p_1}), 
the resolution of the azimuthal angle of the pair is: 
\begin{equation}
\sigma_{\phi} = 
 \left( \gfrac{E}{p_1}\right)^{-\frac{3}{4}} 
 \left[ x_+^{-\frac{3}{4}} \oplus (1-x_+)^{-\frac{3}{4}} \right] \gfrac{E}{E_0}.
\label{eq:sigmaPhi-0-thin}
\end{equation}

A measurement with a good resolution, $\sigma_{\phi} < \sigma_c$, is
therefore possible up to a critical energy, $E < E_L$:
\begin{equation} 
E_L =
\gfrac{\sigma_{c}^4}{2^5} ~ \gfrac{E_0^4}{p_1^3}. ~ 
\label{eq:E_L-0-thin}
\end{equation}

The variation of $E_L$ with the detector density, $\rho$, is shown
in Fig. \ref{fig:E_L}.
In a dense TPC such as a liquid or solid\footnote{Liquid neon does not
allow electrons to drift, but solid neon does \cite{Brisson:1983qb}.}
polarimetry using nuclear conversion is hopeless, since the value is
smaller than 1 MeV.
But, in contrast with conventional belief, polarimetry using nuclear
conversion should be possible in the energy range of interest with a
high-precision gas TPC using an optimal track fitting.

\paragraph{Simulation} 

\begin{figure}
\begin{center}
\iftoggle{couleur}{
 \includegraphics[width=0.49\linewidth]{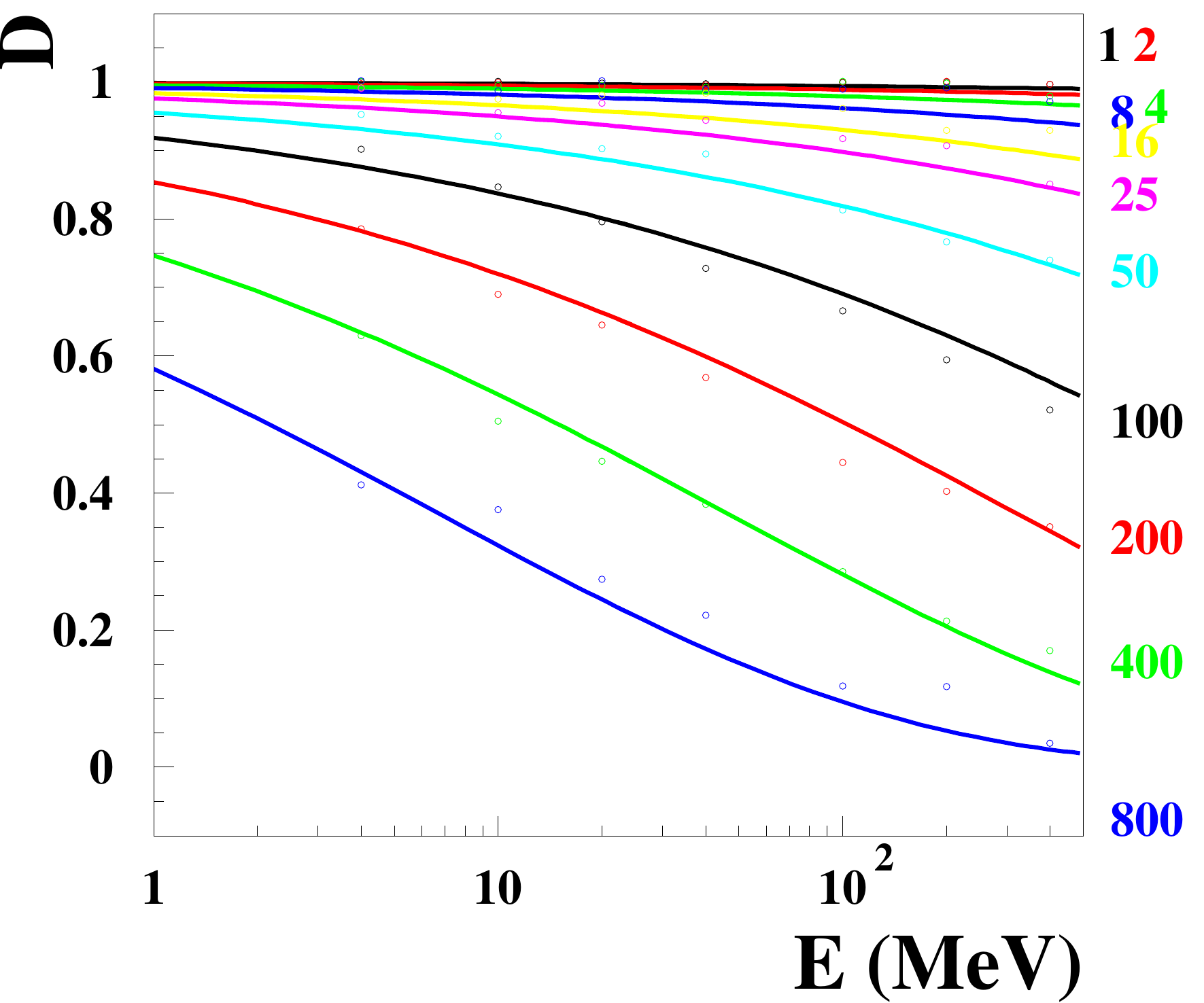}
 \includegraphics[width=0.49\linewidth]{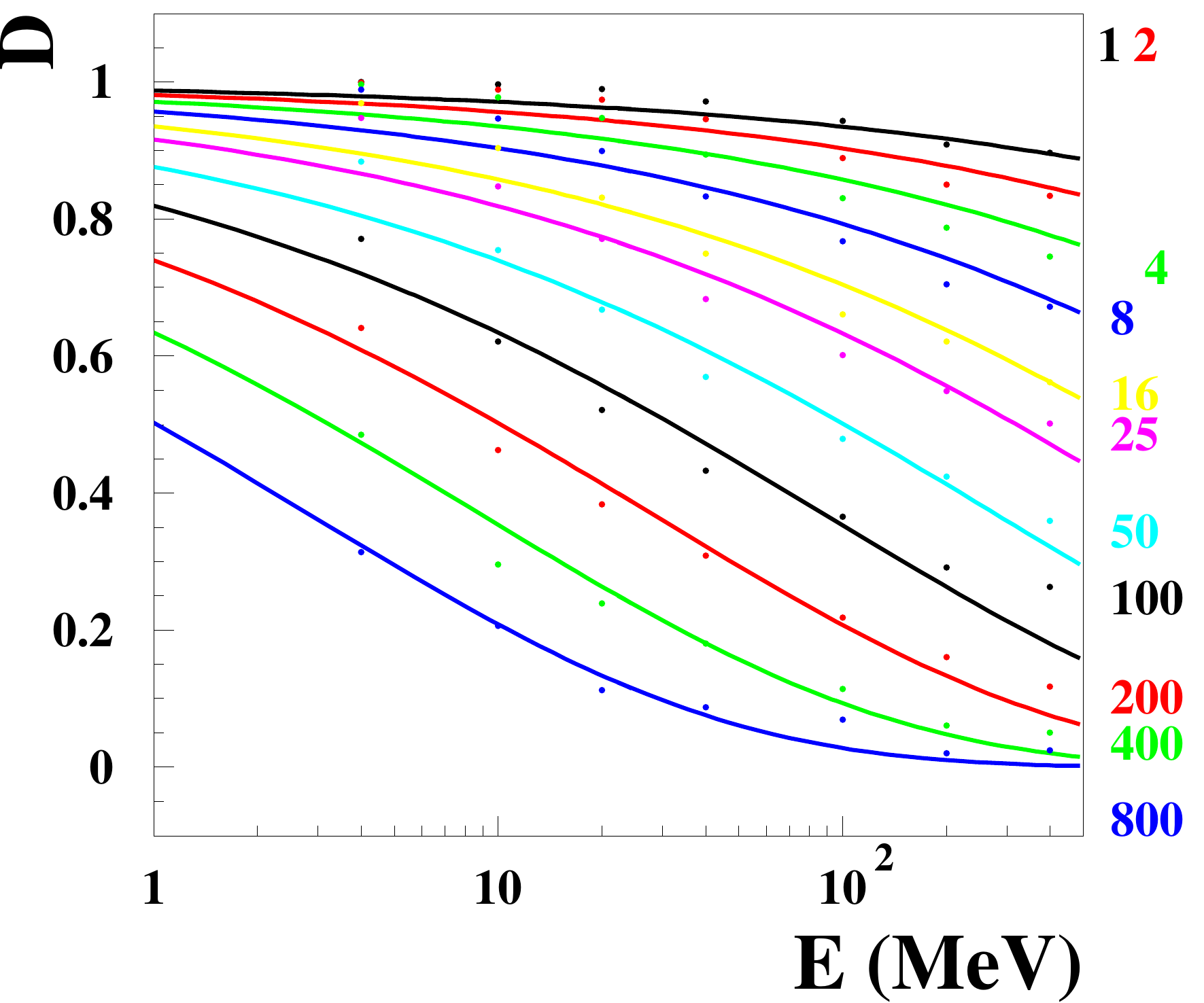}
}{
}
\caption{Energy variation of the dilution $D$ of the measurement of
 $P$, for various values of the parameter $p_1$ that parametrize the
 single-track angular resolution in a thin detector (the values of
 $p_1$ (keV) are listed on the right of each plot. 
Left: 1D weight; Right: 5D weight.
The curves correspond to eq.~(\ref{eq:dil}).
\label{fig:plot:dilution}
}
\end{center}
\end{figure}

We use the simulation to determine the value of the dilution, $D$, as a
function of photon energy and of the detector parameter $p_1$.
Again, the dilution is stronger ($D$ smaller) when $P$ is
measured with the 5D weight (Fig. \ref{fig:plot:dilution}, right) than
for the 1D weight (Fig. \ref{fig:plot:dilution} left).
The data are well fitted by a function 
\begin{equation} 
D(E, p_1) = \exp{ [ -2 (a p_1^b E^c)^2 ] }, 
\label{eq:dil}
\end{equation}
with coefficients $a, b$ and $c$ given in Table \ref{tab:dil}.
\begin{table}[htp]
\begin{center}
\caption{Coefficients of the expression, eq.~(\ref{eq:dil}) of the
 dilution as a function of $p_1$ and $E$.
\label{tab:dil}
}
\begin{tabular}{l|lllll}
weight & $a$ & $b$ & $c$ \\
\hline
5D & $-0.626$ & 0.297 & 0.179 \\
1D & $-0.575$ & 0.445 & 0.159 \\
\end{tabular}
\end{center}
\end{table}

\subsubsection{Precision of the measurement of $P$}

In the general case, the precision of the measurement of $P$ is given
by eq.~(\ref{eq:sigmaP}). 
The number of events is 
\begin{equation} 
N = \int T \eta \epsilon A_{\eff}(E) \frac{\dd N}{\dd E} \dd E
\label{eq:number:evts}
\end{equation}

where $A_{\eff}(E)$ is the effective area, 
$T$ and $\epsilon$ the exposure duration
and the efficiency. 
The exposure fraction, $\eta$, is the ``equivalent on-axis exposure duration''
divided by the effective exposure duration
\cite{Fermi-LAT:2011iqa}. 
For the Fermi LAT, for example, with a equivalent on-axis exposure
duration of $3.2 \pm 1.0\,\mega\second$ over an effective exposure
duration of $29\,\mega\second$, $\eta = 0.11 \pm 0.03$, depending on
the source location and on the telescope configuration (See Fig. 3 of
Ref. \cite{Fermi-LAT:2011iqa}).
In the case of a stand-alone detector such as a TPC, the exposure
fraction is expected to be larger than that of a two-component
telescope like the LAT (a tracker and a calorimeter), and therefore we
can assume that $0.5 > \eta > 0.1$ for steady sources.
The gamma-ray flux for a Crab-like source in the MeV-GeV energy range is 
\begin{equation}
 \frac{\dd N}{\dd E} \approx \frac{F}{E^2}
\end{equation}

with $F \approx F_0 = 10^{-3}\,\mega \electronvolt /(\centi\meter^2 \second)$
\cite{Crab}.
\begin{itemize}
\item
We obtain
a naive estimate of the precision from
eq.~(\ref{eq:sigmaP}), with $N$ from eq.~(\ref{eq:number:evts}) and
$\calA$ equal to a constant value of 0.25.
The results are displayed in Table \ref{tab:naive} as $N$ and $\sigma_{P,0}$.
As expected, the number of conversion events increases with $Z$ for
nuclear conversion for a given detector mass, while the number of
triplet conversions is nearly independent of $Z$, the number of electrons
per unit detector mass being (almost) a constant of Nature.
\begin{table*}[htp] 
 \small
\begin{center}
\caption{Number of conversions observed and $P$ precision, 
 for a Crab-like source and an $M = M_0=1\,\kilo\gram$ detector 
in space for $T=T_0 = 1$ year, with efficiency $\epsilon = 1$ 
and exposure fraction $\eta = 1$. 
No dilution due to multiple scattering is taken into account at this point.
The average asymmetry and $P$ precision are given for the 1D and 5D
weights.
\label{tab:naive}}
\begin{tabular}{lllll|llll}
\hline
\hline
 & $N$ & $N$ & $\sigma_{P,0}$ & $\sigma_{P,0}$ & \calA 1D & \calA 5D & $\sigma_{P}$ 1D & $\sigma_{P}$ 5D \\
 & nuclear & triplet & nuclear & triplet & nuclear & nuclear & nuclear & nuclear \\
\hline
Ne & 59911. & 3237. & 0.023 & 0.099 & 0.188 & 0.228 & 0.030 & 0.011 \\ 
Ar & 97310. & 2918. & 0.018 & 0.105 & 0.188 & 0.229 & 0.024 & 0.009 \\ 
Xe & 261001. & 2586. & 0.011 & 0.111 & 0.192 & 0.235 & 0.014 & 0.005 
\end{tabular}
\end{center}
\end{table*}

For other configurations, the precision is readily
obtained from $\sigma_{P,0}$ by
\begin{equation}
\sigma_P = \sigma_{P,0} \sqrt{ \frac{T_0 F_0 M_0}{\eta \epsilon T F M} }
\end{equation}

\item
A more precise estimate of the asymmetry and of the precision is
obtained from the mean value and the variance of the weights
over the full spectrum.
This calculation involves the 
 average value and RMS of the weight at a given photon energy $E$,
 taken from fits of the data presented in Fig. \ref{fig:1d5d} with
a heuristic function $f(E) = \alpha + \beta / E $, which are
presented in Fig. \ref{fig:1d5d:fit}.
The results are displayed in Table \ref{tab:naive} 
(\calA 1D, \calA 5D, $\sigma_{P}$ 1D, $\sigma_{P}$ 5D).

\begin{figure}
\begin{center}
\iftoggle{couleur}{
 \includegraphics[width=\linewidth]{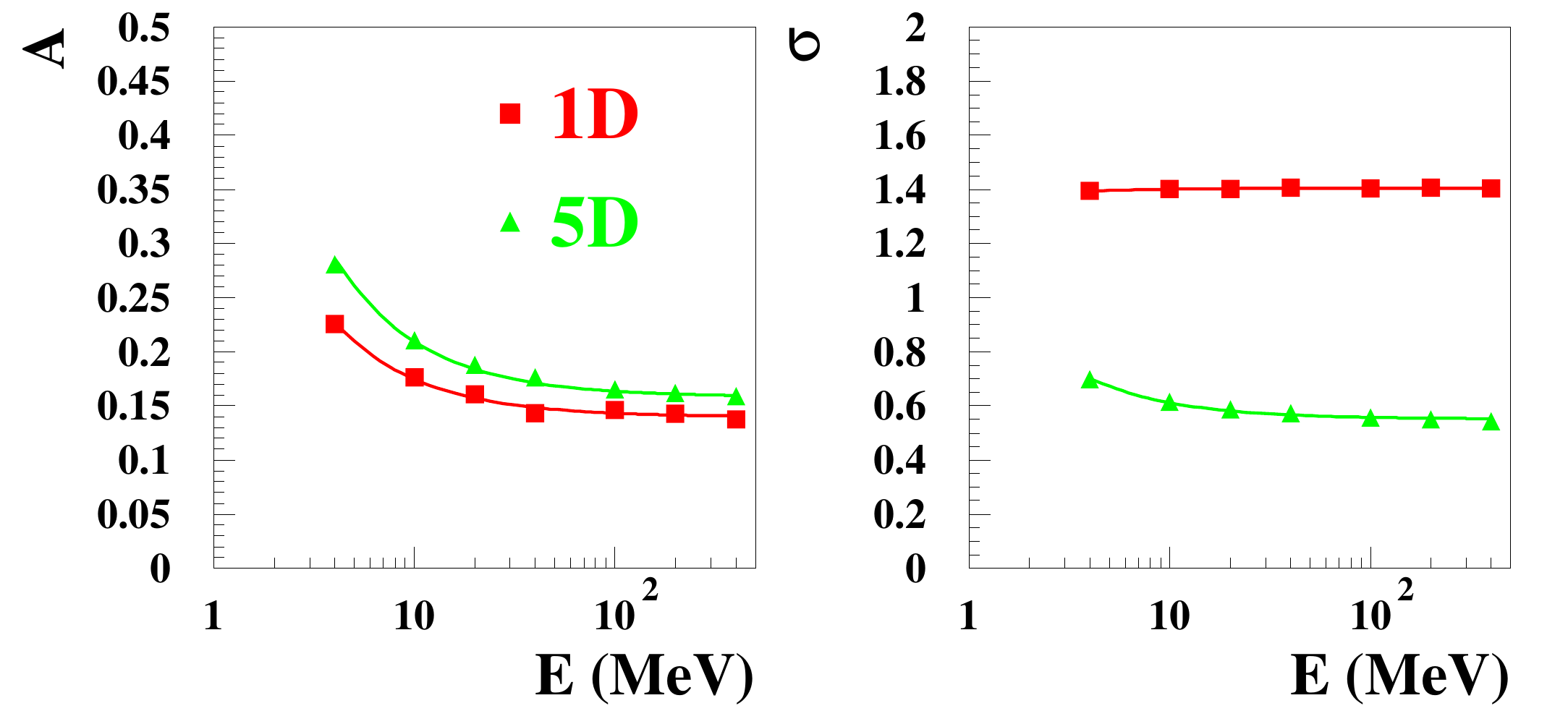}
}{
}
\caption{Fits of the average (left) and RMS (right) values of the polarization asymmetry without dilution, from Fig. \ref{fig:1d5d}, with the heuristic function $f(E) = \alpha + \beta / E $.
\label{fig:1d5d:fit}
}
\end{center}
\end{figure}
\end{itemize}

\subsubsection{Dilution}

In practice, for operation in space, the overall size of the detector
might be a more limiting factor than the detector mass, and one might
want to increase the gas pressure to increase the sensitive mass.
Several factors limit high-pressure operation; micro-pattern gas
detectors show an operational gain that decreases at high density;
even though the mass of the container is a pressure-independent 
fraction\footnote{36 \% for a spherical titanium container of argon
 gas, at the limit of elasticity.}
of the mass of the contained gas, building a low-mass high-pressure container
for a large volume detector will doubtlessly be an issue.
Here we examine the relation between detector density and polarization
dilution. 
For a given sensitive volume, $V = 1\,\meter^3$, we compute the number
and spectrum of converted photons (Fig. \ref{fig:energy-range}).
The dilution factor is taken from eq.~(\ref{eq:dil}), in which $p_1$
depends on the pressure $P_r$ through $X_0$ (see eq.~(\ref{eq:p_1})).

\begin{figure}
\begin{center}
\iftoggle{couleur}{
 \includegraphics[width=0.8\linewidth]{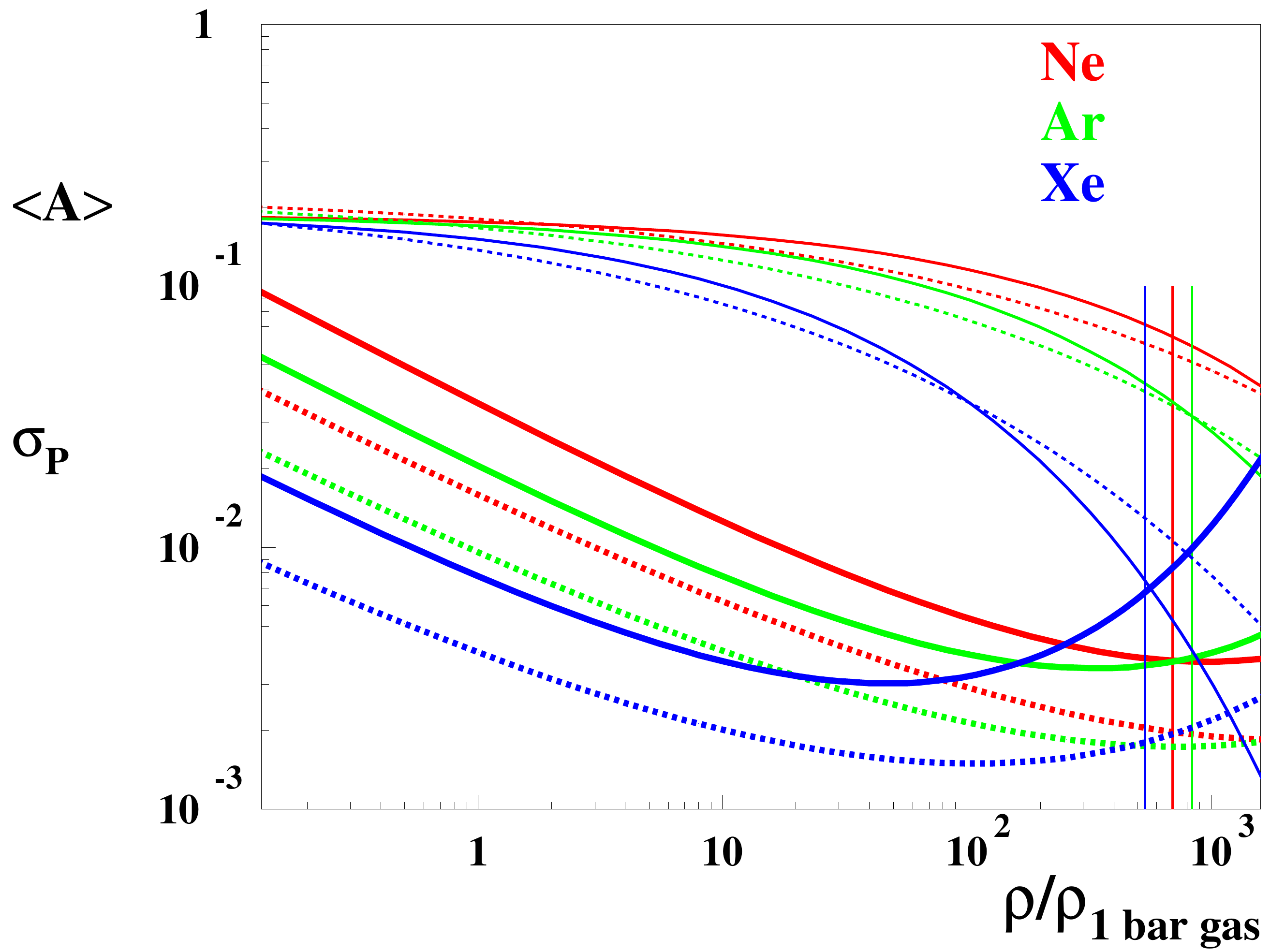}
}{
}
\caption{
Average polarization asymmetry (thin line) and polarization fraction
precision (thick line) as a function of detector density normalized
to the 1\,bar gas density, for a $1\,\meter^3$ sensitive volume detector
exposed for 1\,year and a Crab-like source 
(nuclear conversion, $\eta = \epsilon = 1$).
1D (solid line) and 5D (dashed line) weight.
The vertical lines show the density of the liquid phase.
\label{fig:pressure}
}
\end{center}
\end{figure}

The average asymmetry and the precision for the whole spectrum are
plotted as a function of detector density, normalized to that of gas
at 1\,\bbar, in Fig. \ref{fig:pressure}.
For the density range available to gas detectors, the increase in
statistics overruns the degradation of the dilution, and the
performance of the detector improves with pressure.
It's only at an even larger density, close to that of the dense
(liquid or solid) phase, that the two effects compensate.

Performing polarimetry with a liquid TPC is most likely extremely
difficult in practice. 
Not only does the sensitivity rely on the events with a large opening
angle (Fig. \ref{fig:E_L}) but also on the lowest part of the photon
spectrum (Fig. \ref{fig:plot:dilution}).
At these energies, triggering in the presence of background may be
difficult and extracting the signal of a given source from the
irreducible photon background could also be a problem.

A configuration that is easier to handle is a $5\,\bbar$ argon TPC with the
nominal parameters used in this study ($T = 1$ year, $V = 1\,\meter^3$,
$\sigma = l = 0.1\,\centi\meter$, $\eta = \epsilon = 1$).
The conversion of a 10 MeV photon in $5\,\bbar$ argon is shown in Fig. 
\ref{fig:image10MeV}.
The precision of the (1D) measurement of $P$ for this setup would be
about 1\,\%, with an effective polarization asymmetry of about 15\,\%.

\subsubsection{Background}

$\gamma$-ray telescopes are affected by a huge background.
The dominating contributions are from (charged) cosmic rays that
traverse the detector and from the albedo photons that enter the
detector from below and pair convert into it.
These should be easily rejected, thanks to the excellent pattern
recognition capability of a TPC.
$\gamma$-ray Compton interactions in the detector leave single tracks
that are easily rejected too.
The production of irreducible photons by the interaction of a cosmic
ray into the vessel material, outside of the detector and of the
coverage of a cosmic ray veto, yields photons that enter into the
detector from below, which are rejected.
In the following, we consider that the background is dominated by the
galactic gamma emission.

In the presence of background, the expression for the precision of the
measurement of $P$ eq.~(\ref{eq:sigmaP}) becomes
$\sigma_{P}^\prime = \sigma_w / (\calA^\prime\sqrt{S+B})$,
where $S$ and $B$ are the number of signal and background events,
respectively.
The measured polarization asymmetry, 
$\calA^\prime$,
computed using a weight is 
$\calA^\prime = \calA \times S / (S+B)$,
assuming a non-polarized background with $ \sigma_w$
 unchanged, so that
\begin{equation}
\sigma_{P}^\prime = \sigma_{P} \sqrt{\frac{S+B}{S}}.
\label{eq:sigmaP:bkg}
\end{equation}

The degradation of the performance of the polarimeter is sizable when
$B \ge S$.
In the energy range of interest here, the background in the galactic
plane amounts to 
$f \approx 10^{-2}\,\mega\electronvolt/(\centi\meter^2 \second ~ \steradian)$
\cite{Bouchet:2011fn},
so that $B=S$ when summing over a solid
angle $\Delta \Omega = F / f \approx 0.1 ~ \steradian$, 
corresponding to an apex angle 
$\theta = \sqrt{\Delta \Omega/\pi} \approx 10 \degree$.
The background is not expected to be too much of an issue, except at
the lowest energies at which, for a gaseous TPC, the photon-resolution
angle of nuclear conversion is dominated by the non observation of the
momentum transferred to the recoiling nucleus \cite{Bernard:2012uf}.
In that context, a $10\degree$ 68\%-containment angle is reached for
$E = 5.6\,\mega\electronvolt$.

\subsection{Towards an actual experiment: Full simulation}

Finally we simulate a full spectrum for a 5\,bar
argon detector.
We first cross validate (not shown) the computation of the asymmetry,
dilution and effective asymmetry that were used in 
Fig. \ref{fig:pressure}.
We then implement the following experimental cuts:
\begin{itemize}
\item A cut on the opening angle, $\theta_{+-} > 0.1\,\radian$, to
 ensure that the tracks are 
sufficiently separated for pattern recognition can be performed;

\item A cut on the reconstructed direction of the photon, to ensure
 background rejection, $\theta_{pair} < 10\,\degree$, as determined
 above;

\item A cut on the (kinetic) energy of each of the exiting leptons, to
 ensure proper track reconstruction 
($E_{+kin}> 0.5\,\mega\electronvolt$, 
 $E_{-kin}> 0.5\,\mega\electronvolt$, 
for which the path length in 5\,bar argon is $\approx 30\,\centi\meter$).
\end{itemize}

\begin{figure*}
\begin{center}
\iftoggle{couleur}{
 \includegraphics[width=0.98\linewidth]{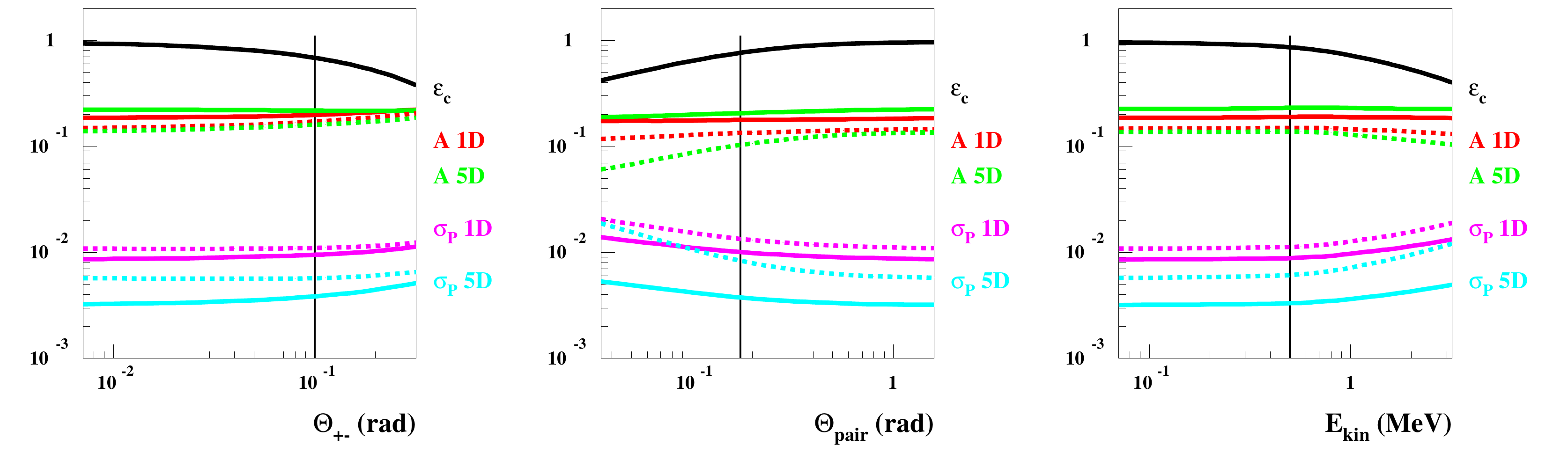}
}{
}
\caption{
Variation of the cut efficiency $\epsilon_c$, of the polarization
asymmetry \calA and of the precision $\sigma_P$ 
 with experimental cuts: from left to right, on the opening angle 
 $\theta_{+-}$, on the pair angle $\theta_{pair}$, and on the lepton 
 (kinetic) energy $E_{kin}$.
With (dashed line) and without (continuous line) dilution due to
multiple scattering.
The cut position mentioned in the text is noted by a vertical line.
\label{fig:cuts}
}
\end{center}
\end{figure*}

Figure \ref{fig:cuts} shows the variation of the cut efficiency
$\epsilon_c$, of the polarization asymmetry \calA and of the precision
$\sigma_P$ with these cuts.
Applying all cuts results in an efficiency of 45 \%, an (1D) asymmetry
of 16.6 \% and a precision of $\sigma_P \approx 1.4 \%$.

\subsubsection{GRB}

For a bright GRB such as GRB 041219A \cite{041219a} the expected
number\footnote{The spectrum of GRB 041219A was kindly provided to us
 by D. Götz up to 100\,MeV, above which we extrapolated it with a
 $\Gamma = 2.06$ powerlaw, for a duration of 120 s.}
of nuclear conversions is $\approx 900$.
After the $\theta_{+-}$ and $E_{kin}$ cuts, 564 events would remain
with an (1D) asymmetry
of 18 \% and a precision of $\sigma_P \approx 33 \%$, and an (5D) asymmetry
of 16 \% and a precision of $\sigma_P \approx 18 \%$.

\subsection{Triplet conversion}

\begin{figure}
\begin{center}
\iftoggle{couleur}{
 \includegraphics[width=0.47\linewidth]{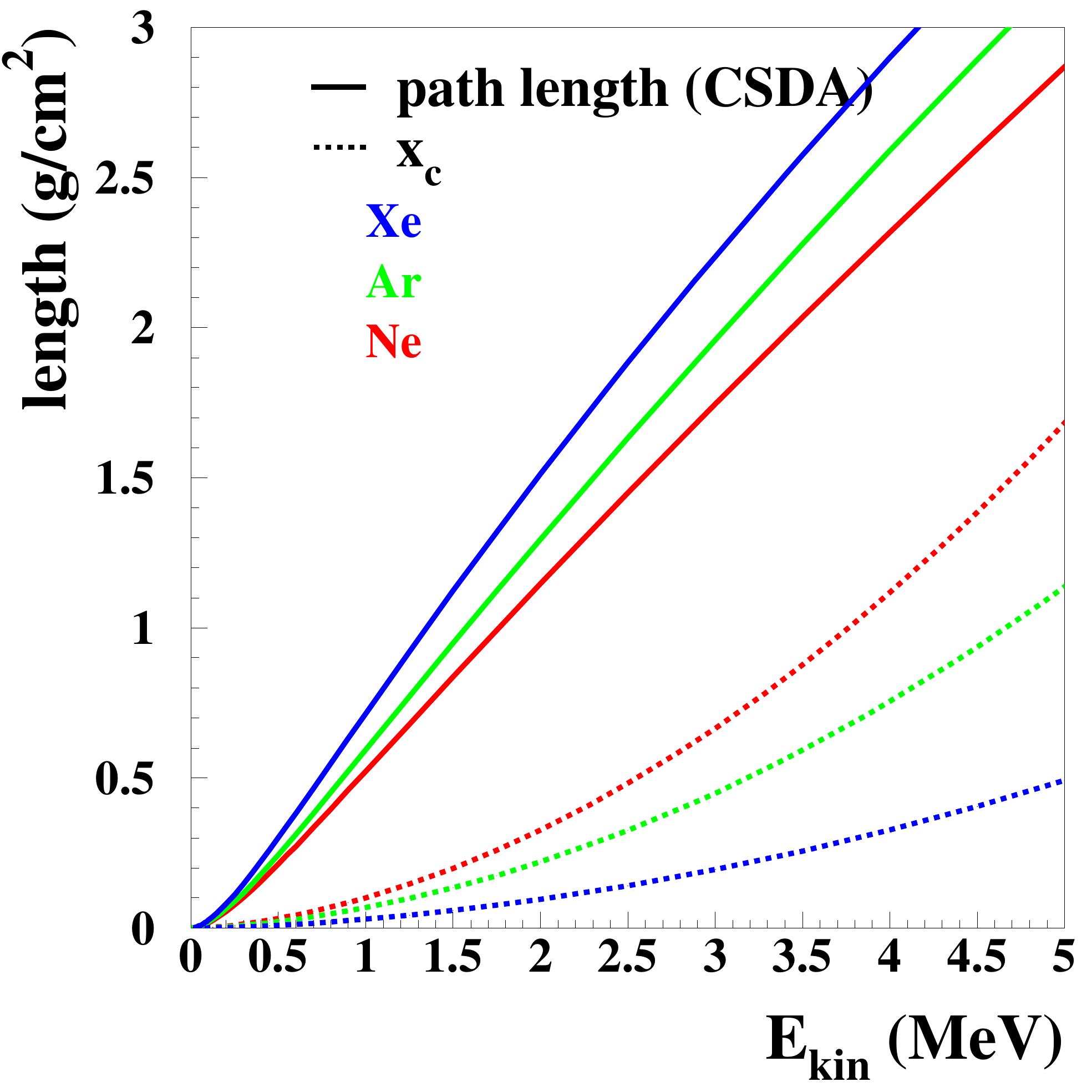}
 \includegraphics[width=0.47\linewidth]{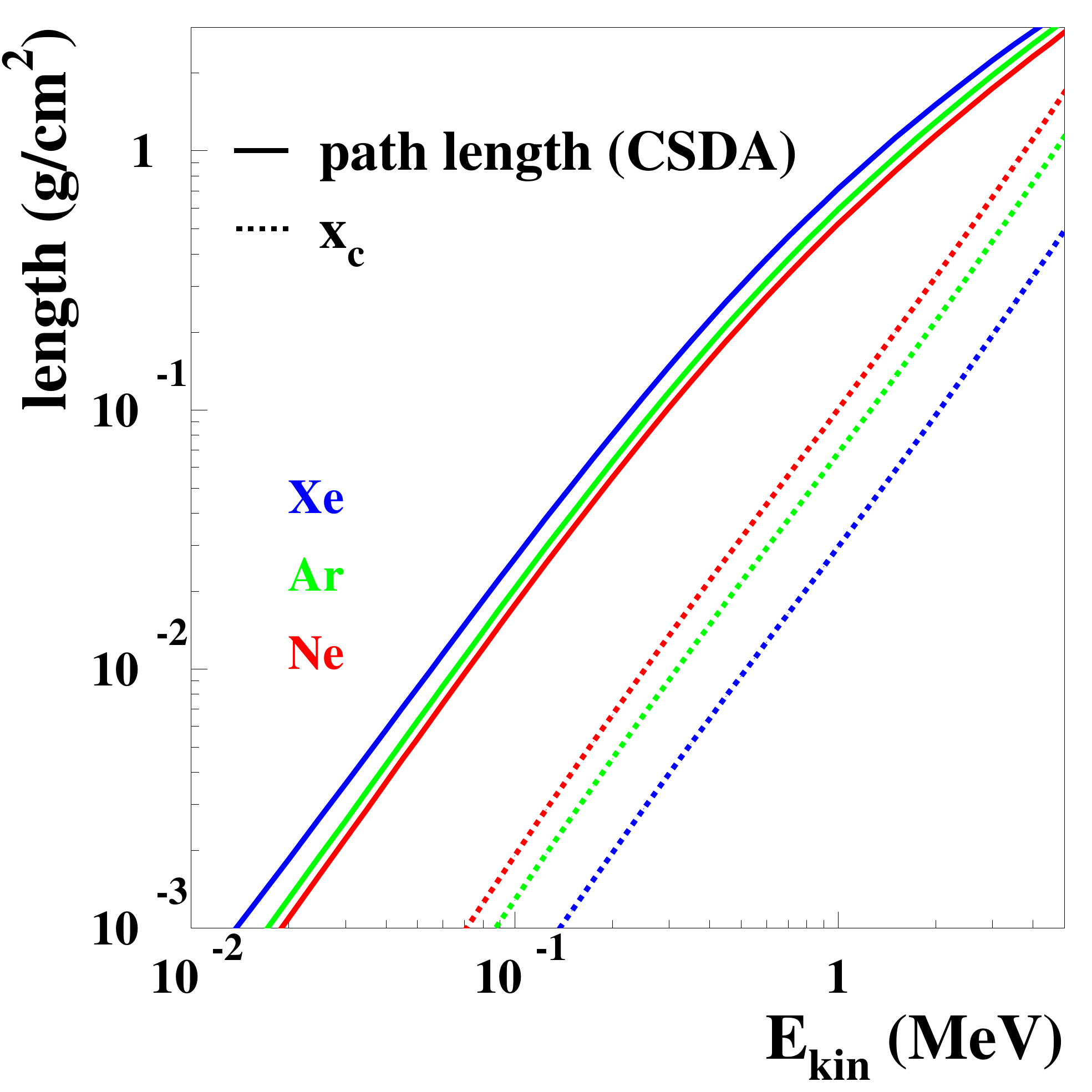}
}{
}
\caption{Variation of the continuous-slowing-down (CSDA) approximation
 pathlength \cite{NIST} with electron kinetic energy (solid line),
 compared to
 that of the path length $x_c$ after which the average deflection
 angle is equal to the critical angle $\sigma_c$ (dashed line)
(left, linear; right, log).
\label{fig:comp}
}
\end{center}
\end{figure}

Only a recoiling electron that has a sufficiently long path length 
inside the detector can be tracked correctly. Also, multiple
scattering affects the measurement of its azimuthal angle. 
In Fig. \ref{fig:comp}, we compare both effects, namely we compare the
path length to the distance $x_c$ after which the average deflection
angle is equal to the critical angle $\sigma_c$ (from
eq. (\ref{eq:theta:MS:PDG})).
In the kinetic-energy range of interest, the limiting effect is
multiple scattering.
This validates a posteriori the constant-momentum, that is the small
energy loss, approximation under which $x_c$ is obtained.

\begin{figure}
\begin{center}
\iftoggle{couleur}{
 \includegraphics[width=0.8\linewidth]{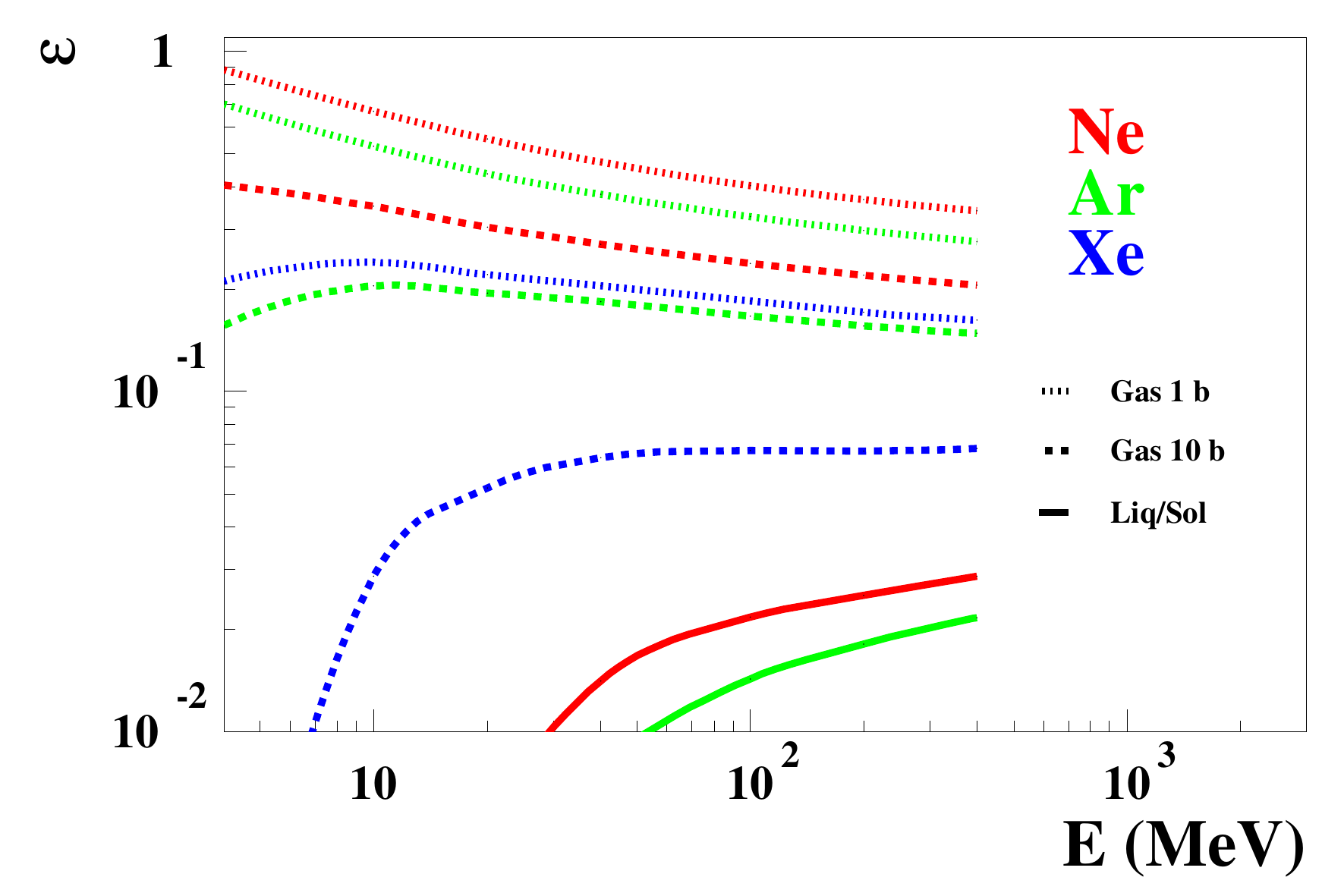}
}{
}
\caption{Fraction of triplet events for which the recoil electron
 undergoes a deflection due to multiple scattering smaller than
 $\sigma_c$ on a path length of 2\,cm.
\label{fig:triplet:fraction}
}
\end{center}
\end{figure}

We can then obtain the fraction of triplet events for which the recoil
electron is measurable, that is for which $x_c$ is larger than a
given value, say 2\,cm (Fig. \ref{fig:triplet:fraction}).
For example in 1 bar Argon, 
$x_c = 2 \centi\meter$ is reached for $E_{kin} \approx 0.17 \mega\electronvolt$,
corresponding to $p_r \approx 0.42 \mega\electronvolt/c$,
that is, for $4\,\mega\electronvolt$ photons, $ \approx 70\,\%$ of the events.
Note that the resolution $\sigma_{\theta t L}$ obtained from the naive
application of eq. (\ref{eq:Vbb:p:2}) yields a value of
$\sigma_{\theta t L} \approx 0.29\,\radian$, smaller than $\sigma_c$ :
after the $x_c$ cut has been applied, we can expect the dilution to
degrade the measurement only slightly.
For triplet conversion, 
increasing the pressure augments the statistics, but degrades the
efficiency of the recoil momentum cut.

\begin{figure}
\begin{center}
\iftoggle{couleur}{
 \includegraphics[width=0.8\linewidth]{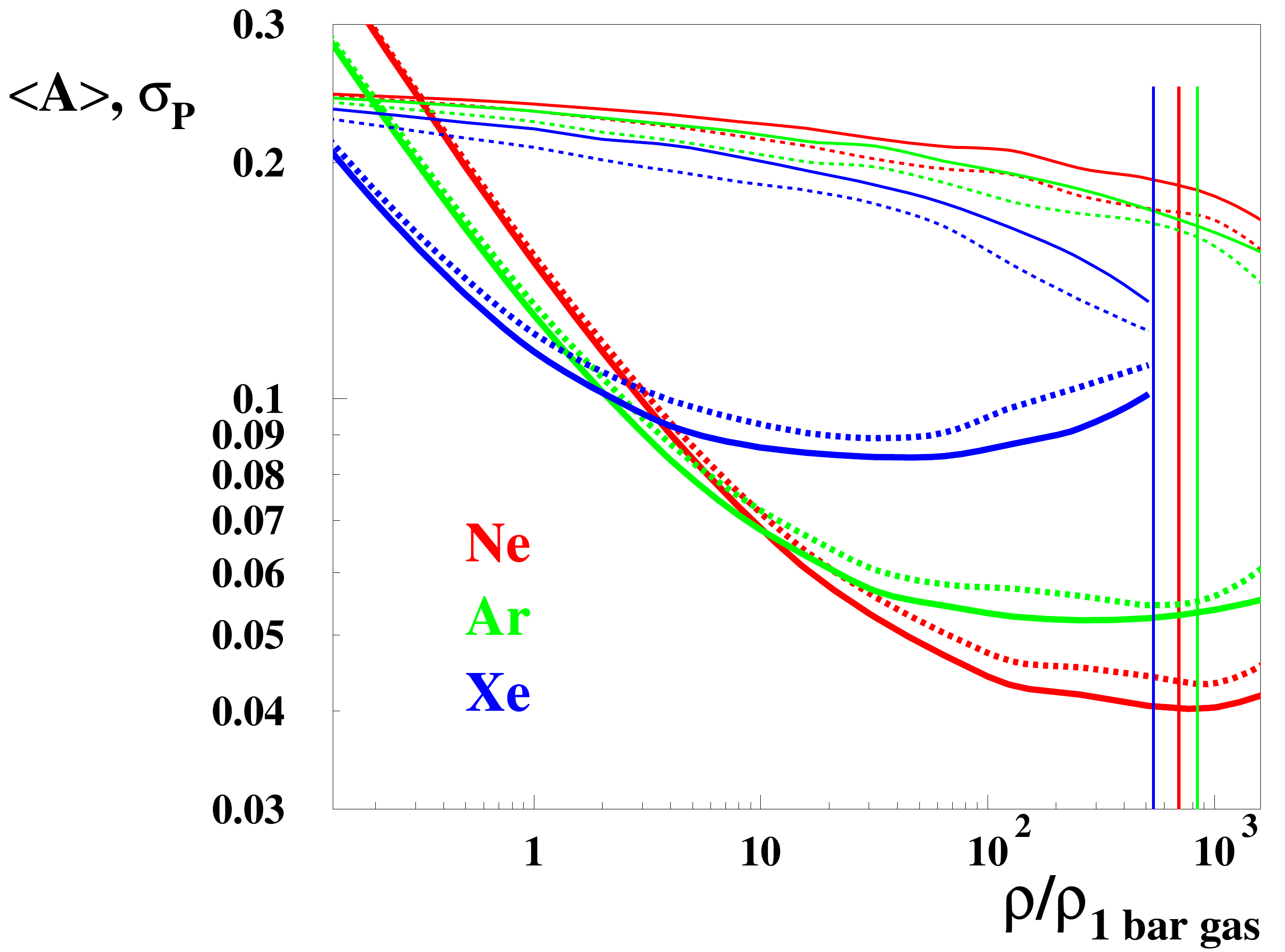}
}{
}
\caption{
Average polarization asymmetry (thin line) and polarization fraction
precision (thick line) as a function of detector density normalized
to the 1\,bar gas density, for a $1\,\meter^3$ sensitive volume
detector exposed for 1\,year and a Crab-like source (triplet
conversion, $\eta = \epsilon = 1$).
Without (solid line) and with (dashed line) dilution.
The vertical lines show the density of the liquid phase.
\label{fig:pressure:triplet}
}
\end{center}
\end{figure}

\begin{figure}
\begin{center}
 \includegraphics[width=0.97\linewidth]{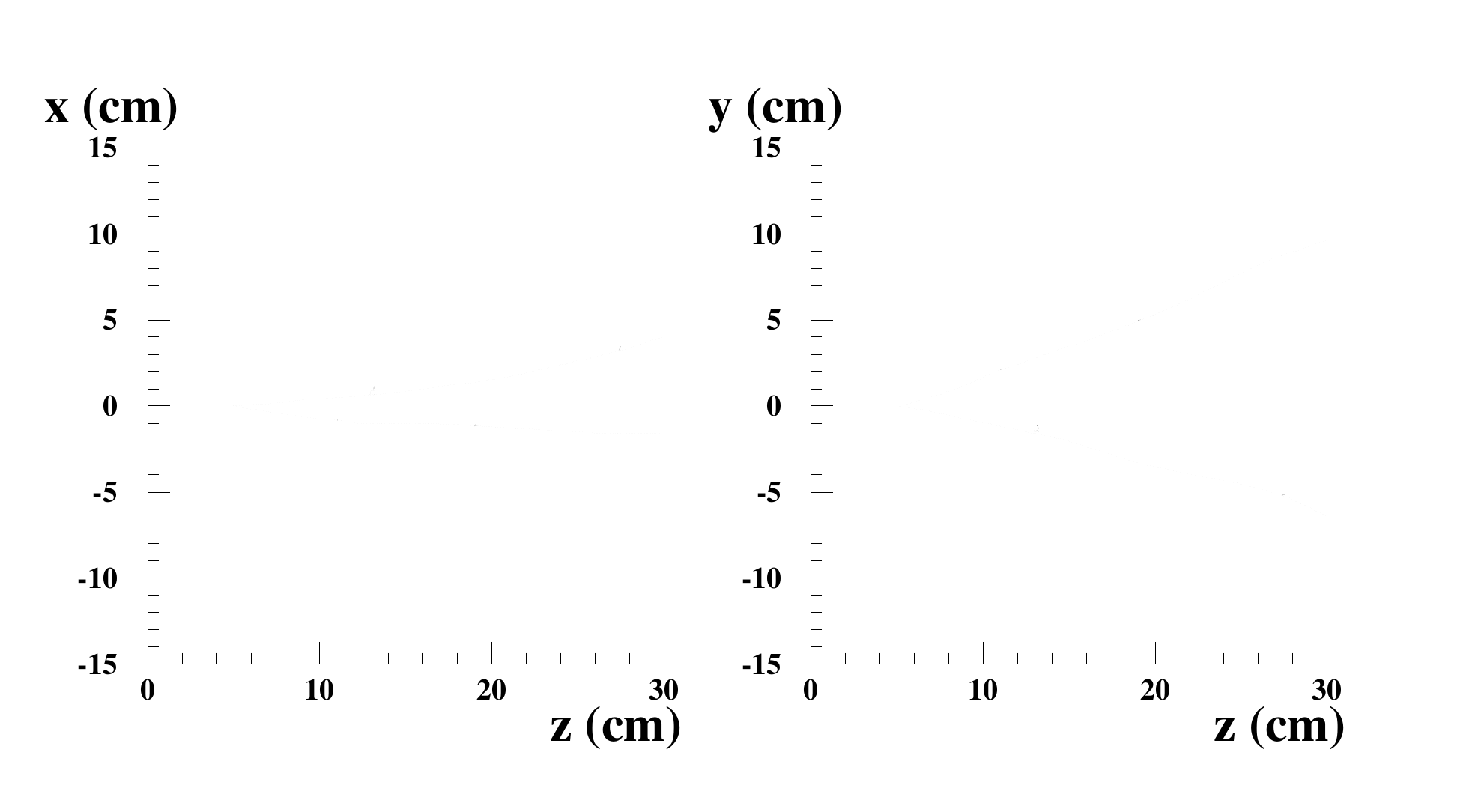}
\caption{$x,z$ and $y,z$ views of a nuclear conversion of a 10\,MeV gamma photon to a 4.3\,MeV positron and a 5.7\,MeV electron in 5 bar argon
(EGS5 simulation\cite{EGS5}).
\label{fig:image10MeV}
}
\end{center}
\end{figure}

The interplay of these effects is estimated with the simulation. The
resolution of the measurement of the azimuthal angle of the recoiling
electron $\phi_r$ is parametrized using eqs. 
 (\ref{eq:Vbb:p:2}) and (\ref{eq:p_1}) and presented in
Fig. \ref{fig:pressure:triplet}.
For 5\,bar argon, we expect a precision of $\approx 8\,\%$ with an
effective asymmetry of 21\,\%.
We note that attempts to increase the statistics and therefore the
precision by using a liquid TPC would provide a small improvement
only.

\section{Conclusion}

We have implemented an event generation, building on the work of Ref. 
\cite{Endo:1992nq}. 
We have interfaced the amplitude calculator HELAS 
with the event generator SPRING. 
For nuclear conversion, the validation showed no difference between
the HELAS pdf and the Bethe-Heitler 
analytical pdf.
Differences between the triplet and nuclear distributions that had
been mentioned previously, 
such as the strong decrease of the most probable value of the opening
angle at low energy for triplet conversion,
or the strong difference between the triplet and nuclear $q$
distributions above a couple of $\mega\electronvolt/c$,
are observed.
Past proposals to increase the polarization sensitivity by a judicious
selection of the events turn out to be inefficient because the possible
improvement is actually counter balanced by the the loss in
statistics.
Only the use of an optimal variable that makes use of all of the
information present in the 5D pdf allows an improvement of the
precision by a factor of about 2.

The dilution of the polarization asymmetry for nuclear conversion due
to multiple scattering in a slab detector is studied with the full 5D
pdf; its variation with slab thickness is found to be quite different
from that which had been computed under the fixed-opening-angle assumption.
The limitations of the converter/tracker remain though, and we turn to
the use of an active target.
In that case, in the multiple-scattering dominated regime that is
relevant in the photon energy range considered here, the momentum
dependence of the single-track angular resolution is parametrized with
a single parameter, a critical momentum, $p_1$, that depends on the
characteristics of the detector:
the dilution is then parametrized as a function $D(E,p_1)$.
We found that with a low density active target, such as a gas TPC, the
sensitivity to polarization of nuclear conversion events is recovered
over most of the photon energy range of interest.

Finally, the simulation of a realistic case, with a $1\,\meter^3$
$5\,\bbar$ argon TPC, a one year exposure to a Crab-like source
(with an exposure fraction of $\eta = 1$)
and with experimental cuts applied yields a $P$ precision of $1.4\,\%$
 (that is, $< 4.4\,\%$ for $\eta > 0.1$).
The precision for GRBs is lower due to the smaller statistics, and we
may only be able to perform GRB polarimetry above 1\,MeV for the
brightest bursts.

~

 {\bf Note added in proof : }
A recent publication \cite{Zhang:2013bna}
 extends previous calculations of the
polarization fraction of the emission of blazars in leptonic and
hadronic jet models up to 500 MeV, that is the gamma energy range
relevant to the present study.
It includes an estimate of the dilution due to the external Compton
radiation (assumed unpolarized) and presents predictions for a sample
of sources.

\section{Acknowledgments}

I would like to express my gratitude to Bogdan Wojtsekhowski (JLab)
and Takashi Kobayashi (KEK) who guided my first steps in polarimetry,
and to my colleagues at LLR for fruitful discussions and constructive
criticisms of preliminary versions of this text.
I also gratefully acknowledge the support by the R\&D program of the
Labex P2IO.


\clearpage
\tableofcontents

\clearpage

\appendix
\section{Supplementary data}

\begin{figure}[hbt]
\centerline{\includegraphics[width=0.65\linewidth]{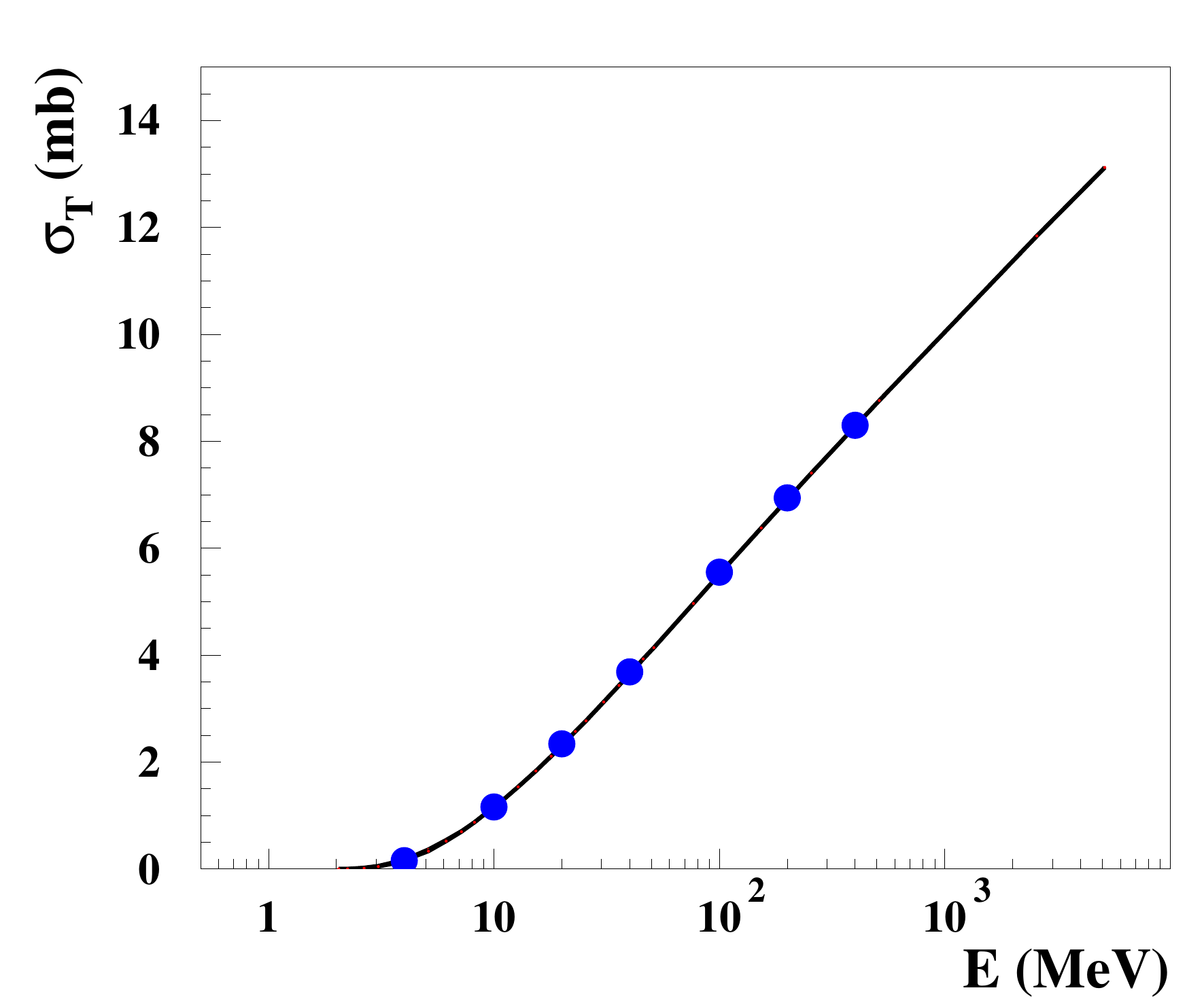}}
\caption{Comparison of the total cross section for triplet conversion (bullets, no screening) with the computation by Mork (solid line)
\cite{Mork1967}.
\label{fig:comparemork}
}
\end{figure}

\begin{figure}[hbt]
 \includegraphics[width=0.49\linewidth]{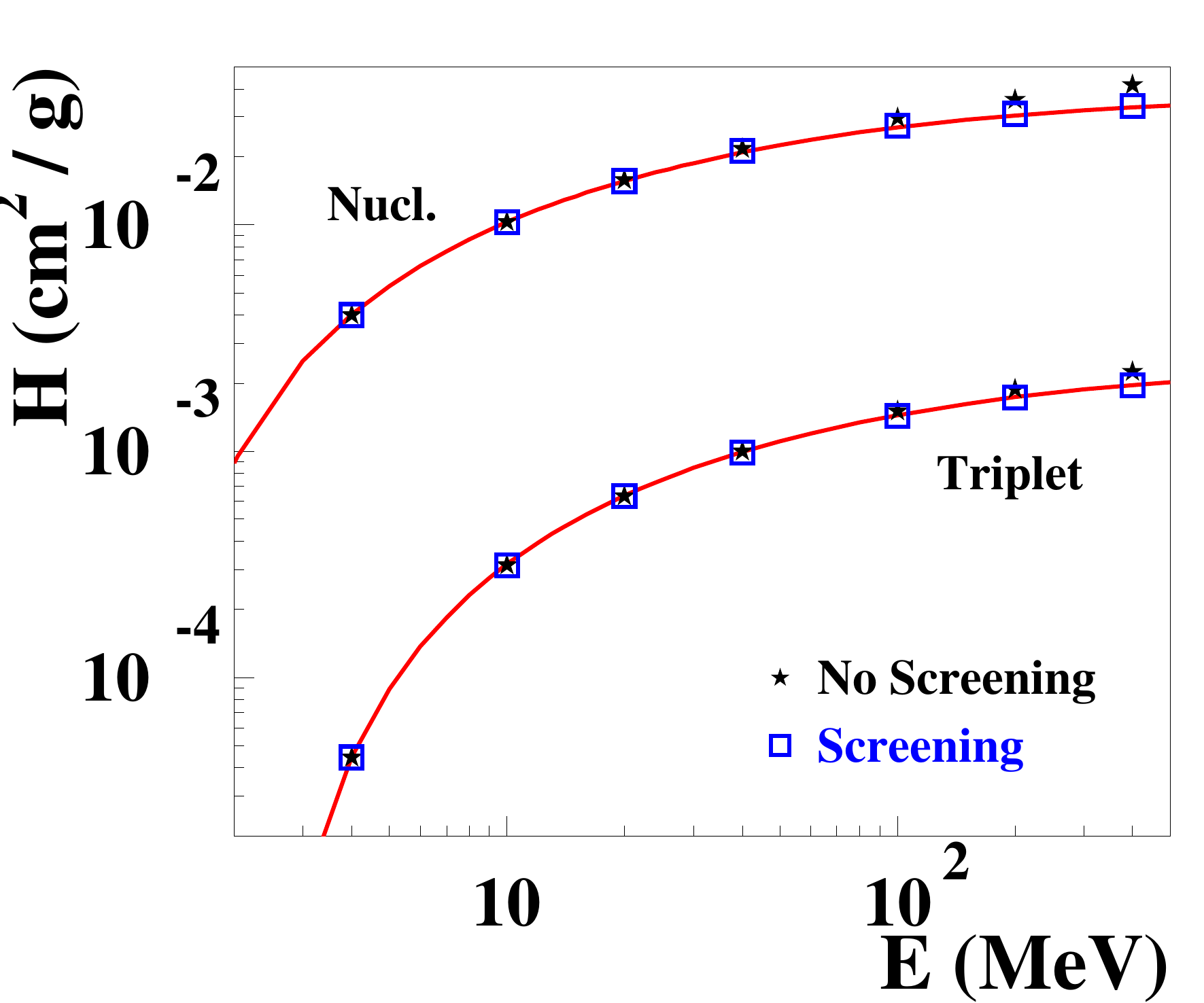}
 \includegraphics[width=0.49\linewidth]{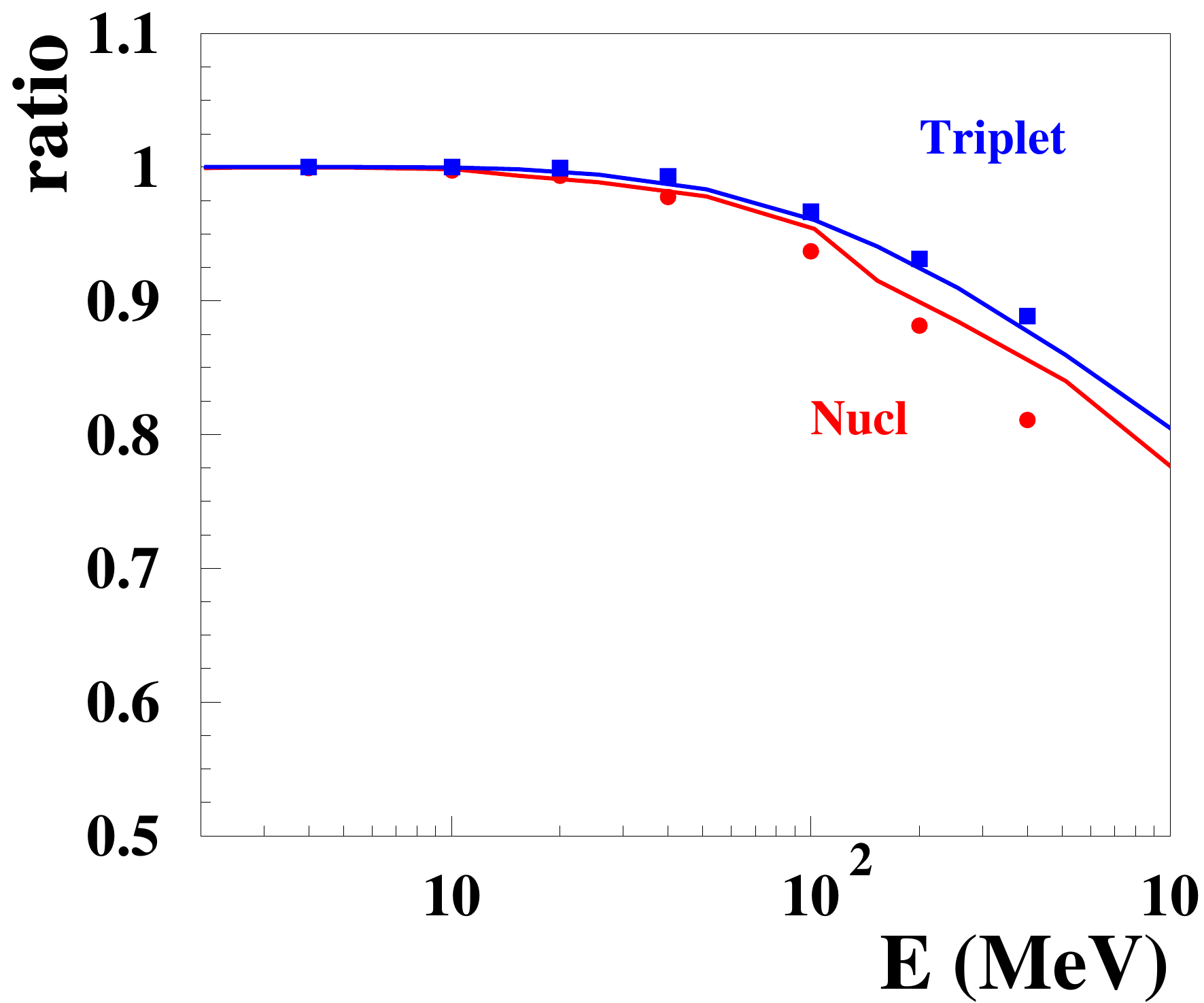}
\caption{Left: 
Comparison of the total mass attenuation coefficients (argon; with
(squares) and without (bullets) screening) with the data from NIST
\cite{NIST} (curves). 
Right: ratio of the cross sections with/without screening 
(symbols) with the results from Ref. \cite{Hubbell1980} (curves).
\label{fig:formfactor}
}
\end{figure}

\end{document}